\documentclass[11pt]{article}
\usepackage[dvips]{graphicx}
\usepackage{amssymb,amsmath}
\usepackage[paper=a4paper,text={16cm,24cm},centering]{geometry}

\newcommand{\be}{\begin{equation}}
\newcommand{\ee}{\end{equation}}
\newcommand{\bel}[1]{\begin{equation}\label{#1}}
\newcommand{\bea}{\begin{eqnarray}}
\newcommand{\eea}{\end{eqnarray}}\date{}
\newcommand{\beal}[1]{\begin{eqnarray}\label{#1}}
\newcommand{\nn}{\nonumber}
\newcommand{\nin}{\noindent}
\newcommand{\nfn}[1]{\renewcommand{\theequation}
           {#1.{\arabic{equation}}}\setcounter{equation}{0}}
\def\d{\partial}
\def\xp{x_{\perp}}
\def\bp{b_{\perp}}

\def\lm{\bar{\lambda} }
\def\T{\hbox to 10pt {\hfill \large $\mathcal \tau$} }
\def\P{\hbox to 10pt {\hfill \large $\cal P$} }
\def\F{\hbox to 10pt {\hfill \large $\cal F$} }
\def\V{\hbox to 10pt {\hfill \large $\cal V$} }
\def\FC{\hbox to 20pt {\large $\cal FC$ } }
\def\Fd{\hbox to 18pt { {\large $\cal F$} \hss {\it~d} } }

\begin{document}

\title{\bf On the parton picture of Froissart asymptotic behavior
}

\author{ {\sc O.V. Kancheli}\thanks{ ~~kancheli@itep.ru} \\
     {\small\it  Institute of Theoretical and Experimental Physics, }  \\
     {\small\it
     117 259 Moscow, Russia. } }
\date{}
\maketitle

 \begin{abstract}
The Froissart  (\F)  asymptotic behavior of high energy
cross-sections, if considered in a parton picture, is  usually
represented  as a kind of behavior that occurs in the process of a
collision of two almost black disks,  filled with partons, when
radii of these \F - disks grow proportional to log's of their
energies.
 ~In this article we briefly review and summarize the main
asymptotic properties of \F - disks, and in particular the
structure of  \F - disks  borders, that one can expect in QCD.~
 Then we consider if it is possible to guarantee the
boost-invariance of transparency $T(s,\bp) = 1 -
\sigma_{in}(s,\bp)$ in a process of collision of two such
\F-disks, where $\sigma_{in}(s,\bp)$ is the total inelastic
cross-section at a definite impact parameter $\bp$.~
 Such a question arises because the mean transverse area of the
overlapping of the colliding \F- disks, at the same impact
parameter $\bp$ and total energy $\sqrt{s}$~,~ varies
significantly with the Lorentz frame, and with it may change the
value of $T$ which is equal to the probability that the particles
do not interact.~~
 We show that it is very hard to make the value of $T(s,\bp)$
boost-invariant at some $\bp$ close to the \F- disk edge for the
case of  saturated \F- disks that one can expect in QCD. This
probably signals that here we have problems with unitarity and
shows that
 the \F type  behavior can be contradictory or the
multiparton system in the high energy Fock wave function must be
arranged in rather specific and unusual way.
 \end{abstract}

\vspace{10mm}
\section*{\bf 1. Introduction}
\nfn{1}

The Froissart (\F) type limitation on the asymptotic behavior of
total cross sections takes place in all local renormalisable field
theories and it was proved on rather general grounds \cite{Froi}.

Not in all such field theories the expected asymptotic behavior of
cross-sections coincides with the Froissart behavior $\sigma_{tot}
(E) \sim (1/m^2)\ln^2 (E/m)$ at $E \rightarrow \infty$, where $E =
m \exp{(Y)}$ is the fast particles energy. For example, in
theories containing only spinor and scalar particles at weak
coupling the total cross-sections decrease asymptotically with
energy.

But in the theories containing vector particles, such as QCD,  we
can have perturbatievly growing cross-sections, that directly
leads to Froissrt behavior. This is especially evident in the
parton picture. Here we have the primary bremsstrahlung-like
contribution to the vector parton spectra of the type
 \bel{pse}
 dn \sim \alpha_s \frac{d \omega}{\omega}
   \frac{d^2 k_{\perp}}{k_{\perp}^2 +\mu^2} ~~,
 \ee
where $\omega,~k_{\perp}$ are partons energy and transverse
momenta, and $\mu$ is some infrared (or confinement) scale,~~
$\alpha_s =g_{QCD}^2/4\pi $. In the first order in $\alpha_s$ we
already have asymptotically constant cross sections.~~ From
(\ref{pse}) we see that this "primary" parton fills the rapidity
interval $\sim 1/\alpha_s$.~ Then with increasing the particles
initial energy $E$ this parton can emit secondary partons, so that
on every step in rapidity of lengths $\sim 1/\alpha_s$ the mean
number of partons in fast particles state duplicates. Such a
cascading transforms the primary parton spectra (\ref{pse}) to the
power growing form
 \bel{psec}
 dn \sim \alpha_s~
      (E/\omega)^{\Delta}~H(\omega,k_{\perp})~
     ~\frac{d \omega}{\omega}
 \frac{d^2 k_{\perp} }{k_{\perp}^2 +\mu^2} ~~,
 \ee
where the parton splitting factor  $\Delta \sim \alpha_s$, and
$H(\omega ,k_{\perp})$ is a soft function at small $k_{\perp}$
 \footnote{
The function $H \sim \exp(-\ln^2 k_{\perp} /\ln(E/\omega))$ at the
large $k_{\perp}$, were we also have taken into account the
variation of $\alpha_s$  with scale.~
 The value of $\Delta$ enters in the intercept
of pomeron trajectory  $\alpha_P(0)= 1+\Delta$~.~
 }.
 The mean number of partons with low energies $\sim \mu$ grows like
$\sim (E/\mu)^{\Delta}$. They are distributed in a transverse disk
area with the diffusion type mean parton density
 $\rho(Y,\xp) \sim \exp ( \Delta Y - \xp^2/r_0^2 \alpha_s Y )$,
where $r_0 \sim 1/\mu ~,~Y = \ln \frac{E}{\mu}$. The mean radius
of this disk $\bar{R}(Y)$ linearly grows with the mean number of
cascading steps
 $\sim r_0 Y \sqrt{\alpha_s \Delta} \sim (\alpha_c/\mu) \ln E/\mu$.
An interaction of such  black Froissart disk ~(~\Fd~) with target
or another fast particle naturally leads to the Froissart behavior
of $\sigma_{tot} (E) \simeq 2\pi \bar{R}^2 \sim
(\alpha_s/\mu)^2\ln^2 (E/\mu )$.

In such a simple picture the very fast particle can be represented
by almost black disk, because the parton density inside the \Fd
continues to grow with $E$, and its transparency $T(Y,\xp) \sim
\exp(-\rho(Y,\xp)\sigma_0)  \rightarrow 0 $. This black disk has
thin grey edge of width $\delta \xp \sim 1/\mu\sqrt{\Delta}$.

In the regge approach to high energy hadron collision one comes to
the same conclusion. Here in QCD we have the BFKL pomeron
\cite{BFKL4} with the intercept $\alpha_P (0) = 1 +
\Delta,~~~\Delta \sim \alpha_s$, which gives the growing
cross-section $\sigma_{tot}(E) \sim (E/\mu)^{\Delta}$. Then, by
the eikonal unitarisation ~(which takes into account the parton
screening and  excludes a multiple count of parton interactions,
leading to a power growth of cross sections)~ we come directly to
the \F type asymptotic behavior of $\sigma_{tot}(E) \sim
(1/\mu^2)\ln^2 (E/\mu)$.

\vspace{3mm}
 But when the parton density inside \Fd becomes high ,
the process of parton gluing can become important even at small
$\alpha_s$, and the soft parton density can stop to grow and to
saturate at values $\sim 1/\alpha_s$. In the regge approach the
transition to the saturation is connected with a contribution of
diagrams with pomeron interactions
 \footnote{And these energies mark the beginning of the energy
region where the regge approach becomes not safe, because here the
average energies on interacting pomerons stop to grow with $E$ and
become not asymptotically large, and the average transverse
distances between neighbor pomerons become small.}.
  The real process of the parton saturation in 4D QCD  is
even more complicated, because the mean transverse momenta of
partons grow with $E$. And the saturation of partons of different
virtualities(transverse momenta) takes place on a different energy
and density scales,  so that  the full parton density continues to
grow, but it decreases from the \Fd center to its edge. We
summarize the main details of this picture in Section 3.

\vspace{3mm}
 The \F type behavior is not so different from that we see  now  in
the scattering data at high accelerator energies. At LHC energies
$\sqrt{s} \sim 7 \div 13$ TeV the transverse profile function
~$\sigma_{el}(Y,\bp)$,  estimated from the behavior of
$d\sigma/dt$ gives the transparency ~$T \leq 0.1$ at impact
parameters $\bp\leq (1 \div 2) GeV^{-1}$. In such a case we
already have in the middle of the fast hadron a clearly seen
embryo of a \F disk. The gaussian-like shape of the edge of such a
''disk'' can also be seen from the the profile functions
corresponding to the high energy data for $d\sigma/dt$.

In the regge approach one can estimate the growth of the \Fd
radius from  the multi-pomeron (\P) contribution to the elastic
amplitude which will dominate the near \Fd edge behavior, where
the contribution of enhanced reggeon diagrams is small. This gives
$r_0 = R(Y)/Y = 2\sqrt{ \alpha'_P \Delta_P}$, where the \P
intercept $\Delta_P$ and the slope $\alpha'_P$ can be found from
the fit of the data at not too high energies and give $r_0 \simeq
0.5~GeV^{-1}$.

The elastic scattering in the \F limit at small transverse momenta
$k_{\bot} \sim 1/R(Y)$ is diffractive. And at large $k_{\bot} \gg
1/R(Y)$ the main contribution to amplitudes comes from the
components of the wave function  without  \Fd and minimal number
of partons in the state (in fact only the valent quarks),~ and it
decreases only power-like in $k_{\bot}$. Such a behavior is
similar to what we already see at TeV energies.
 \vspace{3mm}

The full parton(gluon) density (even saturated ) inside  some part
of \Fd ~(in particular near of the \Fd edge) does not increase
with energy. This means that a low energy particle (or a parton
from other \Fd) moving toward this disk can pass through such part
of disk without interaction with a finite (not decreasing with
$Y$) probability.
 ~This, in particular, is reflected in the fact that $\sigma_{ell} <
\sigma_{tot}/2 $, and the difference $\sigma_{in} - \sigma_{ell}$
~($=$ to the cross-section of the diffraction generation) is
connected with the \F disk transparency distribution.~
 The value of the \Fd transparency can be characterized by the
quantity~
  \bel{tansp}
  T(s,\bp) = 1 - \sigma_{in}(s,\bp) = |S(s,\bp)|^2 ~~~,
 \ee
which gives the probability that the colliding particles (or two
\Fd ) penetrate one through another without any interaction at the
given impact parameter $\bp$. This quantity must be
boost-invariant, that is it should not depend on the coordinate
system in which it is calculated.
 The function  $T(s,\bp)$  is very sensitive to a boost-invariance
breaking ~-~ much more sensitive than other ``big'' cross-sections
like ~$ \sigma_{in} ~~,...$~ themselves !
 The value of $T$ can be directly estimated in the framework of the
quasiclassical approach as the probability that no one pair of
partons will interact. This is especially useful at high densities
of partons.

 \vspace{3mm}

In the regge approach the  $S(s,\bp)$-matrix is boost-invariant by
its construction, but this approach is rather inaccurate for the
description of details of \F asymptotics , because here the
average rapidity on the pomeron lines entering essential reggeon
diagrams becomes small
 \footnote{ But even in the regge approach
the problems with \Fd transparency can be manifested in the
behavior of the diffraction generation cross-sections. ~See
Section 6
 }.

In parton approach we do not meet any problem connected with a
high reggeon density. Here we have the Fock wave function (WF) of
a fast particle given by  the superposition of various parton
configurations,~  so that partons in the dominant components of WF
are arranged in the \Fd like structures. By itself this WF is
evidently not boost invariant - amplitudes of all parton
components of WF change with particles energy $E$.~ Using these WF
we can calculate the cross-sections for various high-energy
collision processes, in particular the $\sigma_{in}(s,b)$ and all
of them must be boost invariant. Their non-invariance will signal
that something is wrong~ (probably the t-channel unitarity of
corresponding amplitudes is violated) .

\vspace{3mm}
 Therefore one can use the requirement of such an invariance of $T
(s,\bp)$ and of other similar quantities as a condition
restricting the properties of the parton system, in our case the
structure of \Fd. In fact, one can expect that the
boost-invariance for all interaction cross-sections in the parton
description can imitate the t-channel unitary conditions
 \footnote{In the parton approach the requirements of the t-channel
unitarity of various amplitudes do not enter explicitly, in
contrast to the regge model.~ At the same time the requirements of
the s-channel unitarity are explicitly encoded in the hermiticity
of the parton interaction hamiltonian. }.

In this paper we consider whether the Froissart type asymptotics
of hadronic cross-sections that naturally arise in the QCD
framework leads to a consistent picture,~  by analyzing the
behavior of $T(s,\bp)$ by the boost-transformation. ~In
particular, we show that the saturated \Fd has the $\delta R(Y)
\sim \sqrt{R/\mu} \sim r_0 \sqrt Y$ wide grey edge. And as a
result for such impact parameters $\bp$ when two \Fd collide with
these grey edges we have problems with boost-invariance of $T$.

 \vspace{3mm}
\nin This article
 \footnote{This article in some points repeats
arguments of the article \cite{Kan1} of present author. }
 is organized as  follows.

In Section 2 we consider a simple example of $(2+1_{\bot})D$
dimensional QCD which is soft and in its high energy behavior can
contain 1-dim saturated  \Fd. In this case the corresponding \Fd
is asymptotically grey and as a result the transparency calculated
in collisions of two \Fd is not boost-invariant.

In   Sections 3 and 4 we briefly summarize the main asymptotic
properties of \F - disks that one can expect in 4D~~QCD~.~ We also
consider in some details the collision of two such \Fd and
describe the characteristic properties of corresponding amplitudes
and cross-sections.

In the  Section 5 we  calculate the transparency in the process of
two \Fd  collisions. We find that it is not boost invariant in
configurations when two \Fd collide with their grey edges.

In the  Section 6 we consider the behavior of various central
diffractive processes with  Froissarons behavior in the regge
approach. It is interesting, because such  processes take place at
the edge of \Fd, where we meat some troubles with the
boost-invariance of $T$ in the parton approach. Here we also find
that the calculated cross-sections violate unitarity, and special
conditions need to be imposed on regge vertexes to avoid such
difficulties.

 Section 7 contains some concluding comments.

In  Appendix we consider in some details the structure and the
fluctuation of the \Fd edge.

\vspace{10mm}
\section*{\bf 2. Froissart type behavior in 2+1 dimensional gauge theory}
\nfn{2}

It is instructive to consider the \F type asymptotic behavior of
cross-sections in $(2+1_{\bot})D$ because here we have only one
transverse direction and the analysis is much simpler. Moreover,~
the QCD theory in 3 space-time dimensions is soft, and the
corresponding BFKL like pomeron is the supercritical regge pole
\cite{BFKL3}. ~Therefore one can expect that here we come to the
Froissart-type behavior

~In $(2+1_{\bot})D$ the Froissart behavior  has the form
 \bel{fr3d}
 \sigma_{tot} (E) ~\sim~  \frac{1}{\mu} ~\ln \frac{E}{\mu}~~,
 \ee
where $E=\mu e^Y$ is the fast particle energy in the lab frame.
 One can expect that such a behavior takes place in $(2+1_{\bot})D$
~QCD  because here the primary gluon emission spectrum ~(over
their energy $\omega$ and transverse momenta $k_{\perp}$)~ is
 \bel{ww3}
  d n \sim \alpha_s \frac{d \omega}{\omega}
   \frac{d k_{\perp}}{k_{\perp}^2 +\mu^2} ~~,
 \ee
where $\mu$ is some infrared confining scale or the effective
gluon mass. From here, taking into account the parton(gluon)
cascading, we come to the parton spectrum
 \bel{pscas}
  d n \sim \alpha_s~  \Big( \frac{E}{\omega} \Big)^{\Delta}
     ~\frac{d \omega}{\omega}
   ~\frac{d k_{\perp}}{k_{\perp}^2+\mu^2} ~=~
 \alpha_s~e^{\Delta (Y-y)}
 ~\frac{dy~d k_{\perp}}{k_{\perp}^2+\mu^2}
~~,~~~~~~y = \ln \frac{\omega}{\mu} ~,~~
   \Delta \sim (\alpha_s/\mu)~~,
 \ee
and therefore to the power growth $\sim (E/\mu)^{\Delta}$ of the
mean number of low energy partons. ~ They fill the  region
 of the transverse coordinate $\xp$  where the mean
parton density $f$ has the form
 \bel{den1}
 f(Y,y,x_{\bot}) ~\sim ~
~\frac{\mu}{\sqrt{Y-y}} ~e^{\Delta (Y - y)}
~e^{ - \xp^2 /r_0^2(Y-y)}~~,~~~~~
  r_0^2 \sim \alpha_s /\mu^3
 \ee
and where $\omega = \mu e^y$ is the parton energy.~ In
(\ref{den1}) the mean parton density $f > \mu$ for $\xp < Y/\mu$,
so this defines the size of the region with small transparency -
i.e. the Froissart disk \Fd
 \footnote{Here we use for the region filled with saturated partons
the same name F-disk $\equiv \Fd$, although it is in this case one
dimensional.}
 with the mean radius $R(Y) \simeq (\alpha/\mu^2) Y$.

In a regge language the behavior (\ref{den1}) corresponds to a
soft BFKL like pomeron \cite{BFKL3} ~with the intercept
$\alpha_P(0) \simeq 1 + \Delta $ ~ and to the reggeized elastic
amplitude
  \bel{pom2}
  A(E,k_{\perp}) ~\sim~
 \Big( \frac{E}{\mu} \Big)^{\Delta -  \alpha'_P k^2_{\perp}  }~~,~~
~~~\alpha'_P  ~\sim ~ r_0^2~~.
 \ee
Then, if we apply the eikonal unitarisation to the amplitude
(\ref{pom2}) ,~ we become the Froissart type  behavior
$\sigma_{tot}(E) = Im A(E,0) \sim (\alpha_s/\mu^2) \ln (E/\mu)$.

The transparency of such a \F  disk when colliding with one low
energy particle (containing few patrons) is ~
 \bel{tran0}
 T_0(Y, \xp) \sim  e^{-\nu(Y, ~\xp)} ~,~~~~
 \nu(Y, ~\xp) ~\sim~
 \sigma_0 f(Y,0,x_{\bot}) ~\sim ~
 ~\frac{1}{\sqrt{Y}} ~e^{\Delta Y - \xp^2 /r_0^2Y}~~,~~
 \sigma_0 \sim \alpha_s /\mu^2~~,
 \ee
where $\nu(Y, \xp)$ is the mean number of collisions of this
particle with  partons of fast \Fd.~ This corresponds to a fully
black \Fd with the thin grey edge whose width $\lambda \sim $ dos
not grow with~$Y$.

\vspace{3mm}
 But, if including  the pomeron interactions, one can expect
that we come to a saturation of the parton density, that almost
stops to grow with energy inside the \F disk
 \footnote{In fact it grows only linearly with Y due to the
contribution of saturated soft partons with higher energies, so
that the transverse density inside saturated \Fd is $f(Y,b) \simeq
(Y-b/r_0) f_0 ~, f_0 \sim 1/\alpha_s$ }.
 In the parton language such a saturation corresponds to the
situation when in parton cascade in rapidity  the locale rate of
parton gluing becomes the same as the local rate of their
splinting. These partons are soft, with average transverse momenta
$\sim \mu$. Their average saturated density far from the \Fd edge
is $f_0 \sim \mu$, and the mean transverse size of $\bar{R}(Y)$ of
the one-dimensional grey F-disk  is $\sim (\alpha_s/\mu^2) Y $
~-~the same as in the unsaturated case, because near the \Fd edge
processes of parton gluing are not essential. But the mean width
of the \Fd edge $\lambda \sim \mu \sqrt{Y}$  will grow with $Y$
for saturated case due to large fluctuations of parton density
near the edge.~~

The transparency of this grey  disk (far from the disk edge) when
colliding with one low energy patron is ~$T_0 \sim \exp (- (
\alpha_s/\mu )^2~ )$.

The shape of elastic amplitude corresponding to such a \Fd at $Y
\gg 1$ as a function of the transverse distance $b=x_{\bot}$ is $
i \tilde{\theta} (\bar{R}(Y) - b)$ where $\tilde{\theta}$ is close
to the $\theta$-function with the smoothed edge~:
 $$
 \tilde{\theta}
(\bar{R}(Y) - b) \simeq  [~1 + \exp{(-(\bar{R}(Y) -
b)/\lambda(Y)}~) ~]^{-1}
 $$

\vspace{10mm}
 \underline{{\bf \emph{Parton density fluctuations in the}}\Fd}
\vspace{3mm}

It could be expected that in every small region inside the \Fd
fluctuations of parton density are Poisson like as in a dense gas.
~~But the behavior of the \Fd edge fluctuations is different. Here
it is reasonable to consider separately the case with or without
parton saturation.

a) Without saturation. ~~~Here the main mechanism which leads to
edge fluctuation motion is connected with  fluctuations at the
first steps of the parton cascade - only such fluctuations are
responsible for variations of \Fd radius in individual events. For
example, if the primary partons are not emitted into the highest
rapidity interval $y_1 < y < Y$, then the following parton
cascading produces  the \Fd with the mean radius $R(y_1)$. ~The
probability of such a fluctuation can be estimated from
(\ref{pscas}) and is
 \bel{}
  w(Y,y_1)  \sim  \exp \Big( -  \nu_1 (Y,y_1) \Big) ~~,~~~~~
  \nu_1 (Y,y_1)  \sim  \int^Y_{y_1} dn
 \sim  \frac{\alpha_s}{\mu} (Y - y_1)~~,
 \ee
where $\nu_1$ is proportional to the number of skipped steps in
the mean parton cascade.
 In particular, it follows from that fact, that a probability of
the bare component of the wave function of a fast particle - that
is the probability that there is no \Fd and the particle contains
only valent components can be estimated as
 \bel{nodisk}
w_0 (Y) \sim \exp(- c Y) ~,~~ c \sim \Delta \sim \alpha_s/\mu
 \ee

b) Saturation. ~~~ Here we see  a wide distribution of the
position in b of the \Fd edge. This position fluctuates from
``event to event''  and defines the variation of the size of the
\Fd.  In the case of one-dimensional saturated \Fd the growth of
$R(Y)$ with rapidity Y in the parton cascade with $n \sim Y$ steps
looks like a random "Brownian motion in $b$ ".~ The number of
steps fluctuates from event to event as ~$\delta n \sim \sqrt{n}$.
This leads to the dispersion of the \Fd edges position  on
 \footnote{
In fact, one should distinguish two transverse scales $r_0 \sim
1/\mu$ connected with the mean distances on which parton moves
when they split and the distance $r_1 \sim  r_0/\alpha_s$ on which
parton density grows from an unsaturated value to a saturated
one.~ Here, for simplification, we do not distinguish these
scales.
 }
  \bel{bordis}
  \lm(Y) ~\sim (\bar{R}(Y)/\mu)^{1/2} \sim r_0 \sqrt{Y}
 \ee
Because the position of disk edge changes in such a random way one
can expect that these fluctuations of the edges position  are
gaussian with the distribution
 \bel{lbd}
w(R,Y) = (1/\lm \sqrt{\pi}) \exp \Big(- \big( R -\bar{R}(Y)
\big)^2/\lm(Y)^2 \Big)
 \ee
and this corresponds

\footnote{ This type form of a smoothed $\theta$-like distribution
for the saturated \Fd was  predicted in \cite{peschan} from a more
detailed consideration.
  }
to such a form
  \beal{tetsm}
 \tilde{f}(Y, b) ~\simeq~  f_0 ~\tilde{\theta} (\bar{R}(Y) - b)
 \equiv  f_0 \int \theta(R-b) ~w(R,Y) ~dR ~=~~~~~~~~~~~~~~~~~ \nn  \\
=~ (f_0 /\sqrt{\pi}) \int^{\infty}_{(b-\bar{R}(Y))/\lambda(Y)}
\exp(-z^2) dz
 ~=~ \frac{f_0}{2} Erfc ~\Big((b-\bar{R}(Y))/\lambda(Y) \Big)
  \eea
of the mean low energy parton distribution  $\tilde{f}$  in the
saturated \Fd.~

The big components of the  Fock wave function of fast particle at
$Y \gg 1$ are given by the superposition of  \Fd states with
different size distributed around $\bar{R}(Y)$.~
 So, the elastic amplitude is the sum
of the grey disks elastic amplitudes with the weight $\sim
\sqrt{w(R)}$. In the inelastic interaction the interference from
components with different \Fd size is small.~ On the contrary, in
the elastic diffractive scattering all components fully interfere.

\vspace{10mm}
  \underline{{\bf \emph{Collision of two saturated} } \Fd}
\vspace{3mm}

At first let us consider processes with large cross-sections.
~When two \Fd with rapidity $y_1$ and $y_2 = Y-y_1$ collide
secondary particles are mainly created from the region of two \Fd
intersections. And this defines the behavior of the total
inclusive cross-section of creation of particles with rapidity
$\simeq y$ which is proportional to the value of the area of this
intersection
$$
\frac{d \sigma (Y,y_1,b)}{dy_1} ~\sim~ \int dB ~\tilde{\theta}
\Bigl( \bar{R}(y_1)-b \Bigr)
 ~\tilde{\theta} \Bigl(\bar{R}(y_2) -b -B \Bigr)~,~~~~Y=y_1 + y_2
$$
at the given rapidity $y_1$ and the impact parameter b and the
full rapidity interval $= Y$.

The total inelastic cross-section $\sigma_{in}(y_1,y_2) \simeq
\bar{R}(y_1) + \bar{R}(y_2) = r_0(y_1+y_2) = r_0 Y + r_0~
O(\sqrt{Y}) $ is boost invariant in the main contribution. The
elastic diffraction cross-section at $Y \gg 1$ for an almost black
\Fd is equal to the inelastic cross section $~\simeq 2
\bar{R}(Y)$.~~ The elastic amplitude is  purely imaginary $ A(Y,b)
~\simeq~ i \tilde{\theta}(r_0 Y - b) $ and universal - it should
not depend on quantum numbers of colliding particles.

The elastic amplitude for a collision of a grey saturated \Fd is
the sum of the grey disk elastic amplitudes with the weight $w(R)$
as in \ref{tetsm}. ~This leads to
 \bel{dsdtsm}
 A(Y, b) \simeq    i (1- T_0) ~\tilde{\theta} (\bar{R}(Y) - b)
 ~~,~~~
~~~~~~~\tilde{A}(Y, k_{\bot}) ~\simeq~
 \frac{2i}{k_{\bot}} (~ e^{ik_{\bot} \bar{R}(Y) - k_{\bot}^2
 \lm^2/4}   - 1 ~)~~
 ~,~~~
 \ee
$$
\tilde{A}(Y, k_{\bot}=0) = 2i (1- T_0) \bar{R}=2 i r_0 Y ~,~~~~~
 \tilde{A}(Y, k_{\bot}) \gg 1/(r_0 \sqrt{Y}) ~) ~
 \simeq -2i/k_{\bot}~~ .~~~~~~~~~~~~~~~~~~~~~~~~
$$
Note that the soft spreading of the \Fd edge leads to a more
smooth (less oscillatory) behavior of $d\sigma_{el}/dk_{\bot}^2$.

The processes of diffraction generation come from configurations
when two colliding \Fd intersect each other with edges. This is so
because only on the edge different components of particles wave
function have  different probabilities to interact.~~ So one can
expect that the corresponding cross section $\sigma_{difg}(Y) \sim
\lm \sim  \sqrt{Y} /\mu $,~ where $\lm(Y)$ is the width of the \Fd
edge in lab frame of one particle.

 \vspace{5mm}
\underline{{\bf \emph{Transparency of}} \Fd {\bf \emph{in} }
$(2+1_{\bot})D$ {\bf \emph{case }}}
 \vspace{3mm}

Let us consider the collision of two such saturated F-disks with
energies $E_1 = \mu \exp{( y_1)}$ and $E_2 = \mu \exp{( y_2)}$
with large full energy, so that $Y = y_1 + y_2 \gg 1$ and at small
impact parameter   $b \sim 1/\mu $. Then it is simple to see that
their mutual transparency is not boost-invariant if considered
only in the states close to the mean one.~

 In the laboratory frame of one of colliding particles  the
transparency $T_{lab} \sim (\tau_0)  \sim \exp (- ( \alpha_s/\mu
)^2~ )$~, because it is determined by the probability of
tunnelling without interaction of only a few $\sim 1$ slow partons
through the finite density saturated \Fd~
 \footnote{Note that only interactions with the low energy
partons from another \Fd are essential, because the local
interaction cross-section decreases $\sim 1/\epsilon$ with partons
relative energy $\epsilon$. }.
 This probability remains finite at $Y \rightarrow
\infty$.

 And in an arbitrary longitudinal frame many partons $ \sim
\nu(y_1) \sim (\alpha_s/\mu) y_1 $ from one disk must tunnel
without interaction trough another disk . As a result, we have
$T(y_1 , y_2) \sim (\tau_0)^{\nu(y_1)} ~\rightarrow 0$ when $y_2
> y_1 \rightarrow \infty$.

Certainly one must  check that there may be not mean components
parton WF , containing the \Fd, but some other more rare parton WF
components, containing small number of partons, which can give the
needed ($\sim const(Y)$) contribution to transparency also in the
center of mass system. of this kind are the states of the fast
particle which contains only small number of partons.
 From (\ref{nodisk}) we see  that the probability of such a
component of the WF of a fast particle is
 \bel{probt}
 w_0(y_i) \sim \exp ( - c y_i)~~,~~~~~~
 y_i \sim Y/2 ,~~~~~~ c \sim 1
 \ee
So, anyway, the contribution to transparency at $b=0$ ~, coming
from this mechanism, is such that
$$
T_{bar}  \sim  w_0(Y/2)  \rightarrow 0~~~ for ~~y_i \rightarrow
\infty~~,
$$
as opposed to $T_{lab}(Y)$  which remains finite at an arbitrary
large $Y$ when parton density in \Fd is saturated (stops to grow
with $E$).~ At $b \neq 0$ the corresponding contributions come
from the colliding states fluctuations in witch  $R(y_1)+R(y_2) <
b$. This gives for $y_1 < y_2$
$$
 T_(y_1, y_2, b) \sim e^{-L \sigma_0 \rho_1 \rho_2 } ~,~~~~
     L = (R_1+R_2-b)\Big( 2R_1 ~\theta(R_2-R_1-b)  +
     (R_1+R_2 - b) ~\theta(b-R_1)  \Big)~~,
$$
where $R_i \sim r_0 y_i$ and $L$ is the length of the two disks
intersection region.~ In cms for $y_1 = y_2$ and $b=0$ this gives
$$
T(y_1, y_1, 0) \sim \exp (- R_1 (y_1)/\sigma) \sim \exp(-c
Y)~~,~~~
 c \sim \alpha_s~~.
$$
Therefore, in $(2+1_{\bot})D$ it is probably impossible to have a
consistent boost-invariant Froissart behavior of ~$T(y_1, y_2,b)$
in the case when the saturation of parton density takes place at
finite value of density  inside of the \Fd.

\vspace{6mm}
  How can this inconsistency be cured ? ~ It seems that there is a
number of possibilities, but they all look rather artificial.

\vspace{2mm} \nin {\bf *}~~
 One possibility is that the parton saturation does not take place
and the mean transverse parton density  does not stabilize at a
fixed value, but continues to grow with energy as $\sim
(E/\mu)^{\Delta_1}$ with some $0 < \Delta_1 \leq \Delta$  and the
transverse parton distribution of type (\ref{den1}).~ In this
case, the transparency in an arbitrary Lorentz system in a mean
state is approximately given by
  \bel{trmod}
T(y_1 , y_2, b) ~\sim~  w_0(y_1) ~w_0(y_2) ~\sim~
  \exp \big(~ - (E_1/\mu)^{\Delta_1}(E_2/\mu)^{\Delta_1}
   \gamma (b)  ~\big)
 \ee
which is boost invariant ~($2 E_1 E_2 \simeq s$)~ and where
$\gamma (b) \sim \exp{-(b/\mu)^2}$. And here for small $b$ the
main contribution to transparency comes from states without \Fd.
~But here the average final state multiplicity of created
particles will grow like  $\sim ~\exp{(\Delta_1 Y)} /Y$~
 \footnote{
Because here we expect that the parton density must saturate due
to parton gluing and whose rate is determined by the square of the
parton density. Therefore, here the only possibility is that
``superfluous'' partons are pushed in the longitudinal direction,~
so that the mean longitudinal size of parton cloud grows with
energy.}.

\vspace{2mm}
 \nin {\bf *} ~~ At high parton densities the parton
system goes into the strong coupling regime near the critical
point with the large density fluctuation in the mean states of
\Fd.~ In this case the transparency even in cms (and $b=0$) can be
large ~($ T \sim 1$).

These large density fluctuations will also manifest in the
behavior of diffractive-generation cross sections which  can
become of the same order in Y as the total cross-section. Note
that for the saturated \Fd such processes are generated only at
the edge of \Fd. But here the entire disk can look like ``edge''.
This type of a soft \F behavior was briefly discussed in
\cite{Kan1}. However, it should be noted that this case seems to
require a very specific fine tuning of the system parameters in
the regge approach. But maybe it can somehow appear in QCD ?

\vspace{2mm} \nin {\bf *} ~~
 It is not excluded that if at very high energies, when in the
$(2+1_{\bot})D$ case one takes into account all high order
corrections in $\alpha_s$, we become the effective $\alpha_P(0)
\le 1$, and then the total cross-sections  decrease ~$\sim
E^{-(1-\alpha_P(0))}$ and the transparency becomes boost
invariant. But this also seems unlikely, especially at small
$\alpha_s$.

\vspace{10mm}
\section*{\bf 3. Parton structure of \F disk in 4D QCD}
\nfn{3}

The main difference of QCD parton spectrum  in $(2+2_{\bot})D$ in
comparison with the $(2+1_{\bot})D$ case, as follows from
comparision of (\ref{pse}) and (\ref{ww3}), is that partons with
high $k_{\bot}$ are also generated  and the weight of the hard
component of \Fd grows with particles energy. In this Section we
beefily summarize the basic properties of such a \Fd that can be
expected in 4D QCD.

 \vspace{6mm}
 \underline{{\bf \emph{Approximate average picture of the }}\Fd}
 \vspace{4mm}

The evolution with $Y$ of the mean transverse parton density
$f(Y,u,\bp) = \d^2 N/\d \bp^2 \d k^2_{\bot}\d Y  $ in the \Fd can
be represented schematically by the non linear generalization
  \beal{nleq}
 \frac{1}{\alpha_s(u)}\frac{\d f(Y,u,\bp)}{\d Y} ~\simeq~ ~
\Phi [f(Y,u,\bp)~]
 +~  c_{0} ~e^{-u}~\mu^{-2}
~\Big[ \frac{\d^2 f(Y,u,\bp)}{\d^2 \bp}\Big]   ~+~
 c_{1} ~  \frac{\d^2 f(Y,u,\bp)}{\d^2 u}   ~+...~~,~~~
 \eea
 $$
 \Phi [f] ~=~ c_2 ~f(Y,u,\bp) ~-~ c_3 ~ f(Y,u,\bp)^2~
    + c_4 ~\alpha_s(u) ~f(Y,u,\bp)^3 ~+...
 $$
of the BFKL-like equation with running $\alpha_s(u) \sim 1/(u +
\alpha_{s0}^{-1} )$,~ where $u = \ln (1 + k^2_{\bot}/\mu^2 ) $ ~is
the partons virtuality, and where $~~\mu \sim r_0^{-1}$ and all
$c_i \sim 1$.~~ Most of the main qualitative properties of \Fd can
be simply found from this equation
 \footnote{We use here such a simplified equation (\ref{nleq})
instated of more precise evolution equation (see \cite{kole}). In
(\ref{nleq}) some gluon properties and effects connected with
gluon coherence are not taken into account explicitly, because
they are not essential for our qualitative picture.~ In equation
(\ref{nleq}) we also included term  $\sim \d^2 f(Y,u,\bp)/\d^2
\bp$ corresponding to parton propagation in the transverse plane
$\bp$. ~This is consistent if parton density is not small.~
 Also this Eq. don't takes into account random fluctuations of the
rate of propagation of various fronts of partons density which can
be especially essential near of the \Fd edge - see footnote 18.
}.~

The mean parton structure of \Fd can be schematically represented
as a system of inserted into each other saturated disks(sub-disks)
with a different virtuality $u$ and rapidity $y,~~Y > y > 0$. ~ At
$y \sim 1$  the soft disk is the largest and its average
transverse size in $\bp$ grows with rapidity as $\bar{R} (Y) = r_0
Y$ with small corrections
 ~\footnote{In fact there can be small corrections
\cite{LRfroi,RisFroi} to the dependence of the average \Fd radius
from particles rapidity, coming from \Fd edge oscillation and
which are not taken into account in Eq. \ref{nleq} , so that $R_F
(Y) = r_0 (Y - c \ln Y)$. See also Appendix. }.
  Note that evolution equation (\ref{nleq}) does not take explicitly
into account the nonperturbative QCD effects which can be
essential for grow of the soft \F sub-disk, especially near the
edge of \Fd where the parton density is low
 \footnote{It seams that the essential in (\ref{nleq})
nonperturbative effects can be simply imitated by  the large value
of QCD coupling $\alpha_s(u) \sim 1/(u + \alpha_{s0}^{-1})$ frozen
at small $u$~ and the effective gluon mass $\sim \mu$. }.

The mechanism by which the distribution of the parton density
$f(Y,u,\bp)$ in \Fd  is generated by the evolution Eq.
(\ref{nleq}) is schematically illustrated by Fig.1  where the
whole triangle corresponds to the region of \Fd filled by  soft
partons (small $u \sim 1$) with various rapidity $0 < y < Y$.
\vspace{20mm}
\begin{figure}[h]  \centering
\begin{picture}(160,155)(80,20)            
\includegraphics[scale=.4]{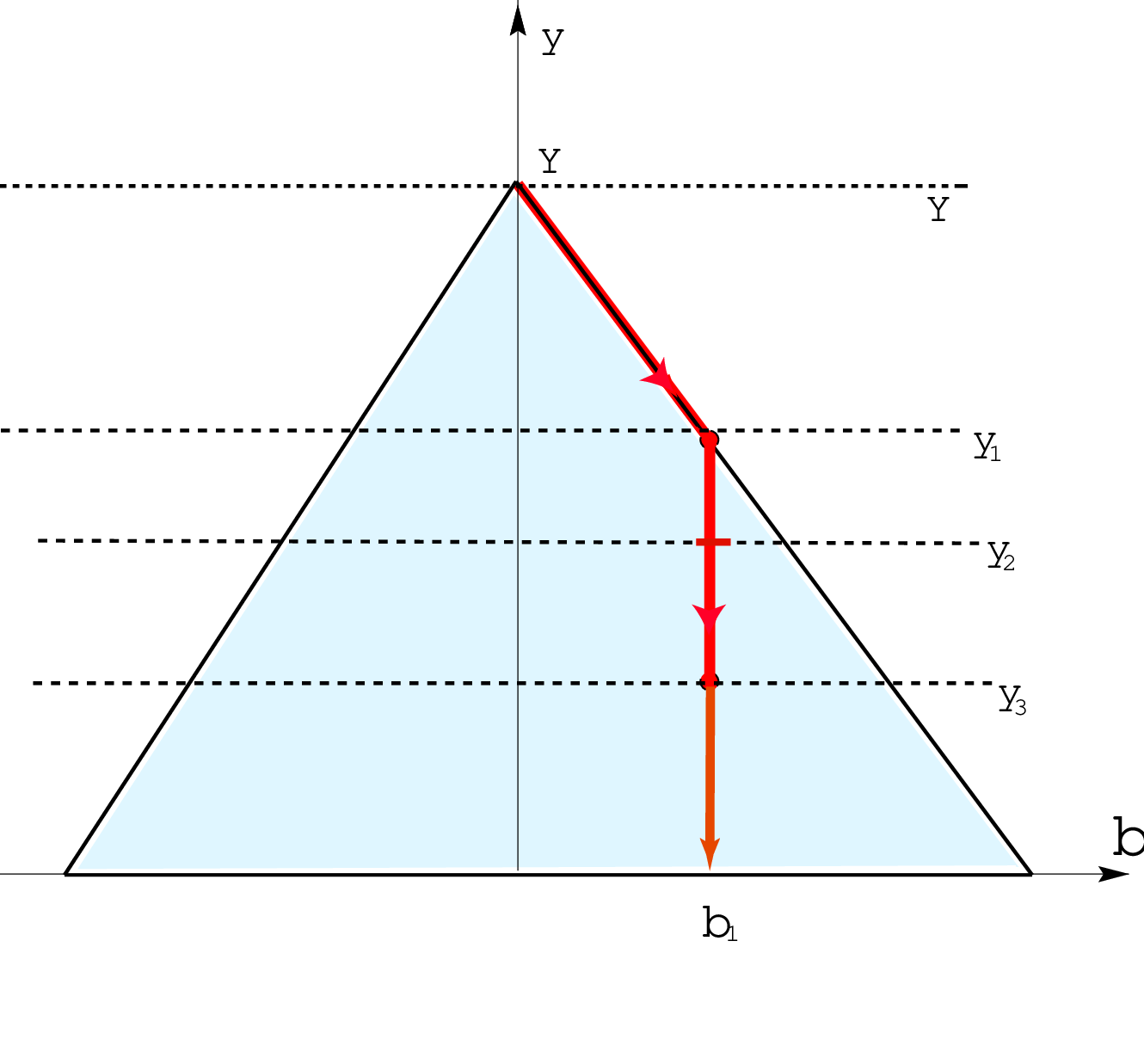}
\end{picture}
\parbox{14cm}{  {\bf Fig.1}~~
The main ``trajectories'' (red lines) of the parton evolution in
$(y,\bp)$ that contributes to the creation of partons with high
virtuality $u \gg 1$ at the definite transverse position $b_1$ in
\Fd. At $y_1 = Y - b_1/r_0$  the parton density $f(Y-y_1,b_1,u)
\simeq O(e^{-u})$.~ At $y_2  \sim y_1 - u^2/c$ the u-parton
density reaches the saturation level $f(Y-y_2,b_1,u) \sim 1$ and
then  remains constant.
 } \label{Fig.1}
\end{figure}
The positions of  more hard sub-disks edges can be marked by the
saturation boundary  -  the line  $u = u_s(y,\bp)$ which divides
regions of \Fd  so that partons with $u < u_s(y,b)$ are in the
saturated state. Parametrically this boundary can be found from
Eq.(\ref{nleq}) by the condition $\Phi[f_s(y,u,\bp)] = 0$, which
gives $f_s(y,u,\bp) \sim 1$ for the value of the saturated
density.

 ~To estimate the growth of $f(y,u,\bp)$ with $Y-y$ one can take into
account that the parton diffusion in the transverse coordinate
$\bp$ is small for large $u$.
 At the given $(u,\bp)$ the parton density, before it reaches the
saturation level, grows with $y_b$ as
$$
f \sim f_{in} \exp{(c y_b /u)}~~,   ~~~y_b = Y -y - \bp/r_0~~,~~
~~f_{in} \sim e^{-u} ~~.
$$
 The value of $y_b $ can be seen at Fig.1 - as the ``length'' of
path in rapidity, which starts from the soft edge of\Fd, and on
which the density of partons with high virtuality $u$ reaches the
saturation level.
 On this path $f$ reaches $f_s \sim 1$ at $y_b \sim u^2$.~
 From here one finds the simple expression for the u-partons
saturation edge
 \bel{satline}
 u_s(y,\bp) ~\sim~ \sqrt{y_b} ~\simeq   ~\sqrt{Y -y - \bp/r_0}
 \ee
This evolution of parton density $f$ with $y$ can also be
considered as a propagation in ``$u$-direction'' (at fixed $\bp$)
of the front of saturation wave described by Eq.(\ref{nleq}) gives
 ~\footnote{This process is similar to growth of \Fd radius
$R(Y) \sim v Y$ and is described by the same type nonlinear
equation. The difference is that in the case of $R(Y)$ the
velocity of disk edge grows~ $v \sim \alpha_s (1)$ does not depend
on $\bp$ and therefore the velocity is almost constant. But for
the growth in the $u$ direction the velocity of edge growth $\sim
\alpha_s(u) \sim 1/u$ depends on $u$~. }
$$
u_s(y)  ~\sim \alpha_s(u_s) ~y_b
       ~\sim u_s^{-1} ~y_b~~,
$$
and this again leads to $u_s(y) ~\sim~ \sqrt{y_b}$~.

\nin It is  used here that the soft disk shrinks linearly as $r_0
y$,~and that the partons with high u almost do not move in the
transverse plane.~ It follows from (\ref{satline}) that the mean
size of the saturated sub-disk with large virtuality $u$ is
 $$
      R(y,u) ~\simeq~ r_0 ~(~ Y - y - u^2/c ~)~.
 $$
Using all this one can represent approximately the mean saturated
parton density in \Fd  in a simple form
  \bel{partde}
f_s (Y,u,\bp) \sim
 ~~\tilde{\theta}  (~ Y - u^2/c - \bp /r_0 ~) ~,~~~
 \ee
where $\tilde{\theta}$ differs from $\theta$-function only near
the \Fd edge.

Such  picture of \Fd  is mostly quasiclassical, because the parton
densities in mean configurations are high. The average number of
partons with virtuality u in \Fd asymptotically grows with Y as
$\sim Y(Y-u/c)^2 e^u $, and the full parton density distribution
$\rho(y,\bp)$ in \Fd is
 \beal{fpd}
 \rho(Y,b) = \int dy~du~ e^u ~f_s (Y,u,\bp) \sim
 ~\exp{\Big(  \hat{c} ~\sqrt{ Y - \bp/r_0 }  \Big) }~,
            ~~~\hat{c} \sim 1 \\
\nn N(Y) = \int d^2 \bp ~\rho(Y,\bp) ~\sim
 ~e^{\hat{c} ~\sqrt{Y}}~~,~~~~~~~~~~
 \eea
where $ N(Y)$ is the mean number of partons in \Fd as a function
of particles rapidity at $Y~\gg~1$.~ One can expect from here that
because the parton density inside the \Fd is large, the
nonperturbative effects are inessential in the internal parts of
\Fd.

 \vspace{8mm}
  \underline{{\bf  \emph{Statistical properties of} } \Fd}
 \vspace{3mm}

Usually we represent a fast particle with its parton cloud as a
complicated pure state. But at $Y \gg 1$  the mean parton number
$N(Y) \sim \exp{( cY )}$ in the \Fd cloud is very big.~ And then
in some cases it can be more adequate to consider this particle
(or big part of it) as a macroscopic object with the finite
entropy $S(Y)$ which is growing with $Y$.

Furthermore, the low energy partons (especially soft with $u \sim
1$), corresponding to last stages of parton cascade are entangled
with vacuum partons(fluctuations), and this also can be considered
as a source of \Fd  entropy  in the parton description of a fast
particle.~~ If interpreted in such a way, the entropy of the large
soft \F sub-disk is $S(Y) \sim N_{soft} \sim (R(Y)/r_0)^2 \sim
Y^2$.
 Partons in more hard \Fd sub-disks, although they are strongly
correlated with the soft \Fd, can also give some additional
contribution to fast particles $S(Y)$.

Even if we forget about the entanglement of low energy partons
with vacuum fluctuation one can estimate the contribution to
$S(Y)$ from the examination of a ``density of levels'' near the
main ``parton state'' of the fast particle  with 4-momentum $P$.
To find the energy distance $\Delta E $ to the nearest state one
can estimate the minimal 4-momentum $k$ of the test particle
colliding with \Fd that can excite the fast particle or create an
additional particle. This corresponds to the condition $(P+k)^2
-P^2 \sim m^2$ which gives~ $2 E ~k_0 \simeq m^2 $ ~for~ $\vec{k}
= 0$ ,~i.e. we become~ $\Delta E \simeq k_0 \sim m ~e^{-Y}$. So,
the entropy can be estimated as $S(Y) \sim \ln (m/\Delta E) \sim
Y$.

To ``discover'' that the isolated fast particle is in a pure state
one needs a large time. This time  grows with the energy for
particles states containing \Fd,  and it is of the order of the
Poincare's recurrence time $t_P \sim m^{-1}\exp{(S(Y)))}$, which
for $Y \gg 1$ can be  bigger than the characteristic ``Compton''
time, that is proportional to the particles energy $m^{-1} \exp{(
Y)}$ , and during which the final state is prepared after the
interaction
 \footnote{If $S(Y) \sim Y^2$ then, for example, at large enough
$E_{Lab} \sim 10^{19} eV$ when $Y \simeq 20$ the value of time
$t_P \sim \mu^{-1}\exp{(S(Y)} \sim \mu^{-1}\exp{(20^2)} \sim
10^{100} s$, and such $t_P$ is much more large than any
macroscopic ``preparation'' time of such a  particle  initial
state. }.

The process of the particles elastic diffraction also operates on
the time scale $\sim m^{-1} \exp( Y)$, that is short relative to
$t_P$. But this process ``measures'' only the part of the wave
function connected with the particles full momentum, while the \Fd
entropy is connected with the ``internal part'' of the wave
function, and all these components diffract in the equal way. The
difference in absorbtion of various components of the wave
function comes mainly from the  edges interaction of \Fd , and it
gives the contribution to the diffraction generation.

 \vspace{10mm}
  \underline{{\bf  \emph{Fluctuations of} } \Fd}
 \vspace{3mm}

The physical state of a fast particle is a complicated
superposition of multiparton states which, on average, can be
considered as an almost black \Fd.~
 One can expect that there are local  Poisson-like fluctuations of
the parton density around  average values in the interior parts of
\Fd, as in every dense gas or liquid. ~But in processes with high
mean multiplicities their contribution to various effects is not
so essential~
 \footnote{In principle, one can not exclude that such a parton
system is in a critical point and in this case all fluctuations in
\Fd are big. Such possibility was discussed in \cite{Kan1}.  But
to have this behavior one needs some special symmetry which is not
seen now. }.

The major global big fluctuations in the parton population of \Fd
come from the fluctuations at early stages of the parton cascade.
For example, if, according to Eq. (\ref{pse}), the primary partons
are not emitted to the rapidity interval $Y > y > y_1$, then the
mean size of \Fd in these components of the parton wave function
is only $r_0 y_1$. The probability of such a fluctuation (it is
the weight of these components of the parton wave function ) is
 \bel{flucy}
\sim~ e^{- (Y - y_1 ) \alpha_s }.
 \ee
That is, the probability that there is no \Fd in the wave function
of fast incoming particle can be estimated as $\sim
\exp{(-\alpha_s Y )}$ ~
 \footnote{ Or more accurately $\sim \exp{(- c Y  \ln Y)}$, if one
takes into account the contribution of primary high $k_{\bot}$
partons and the running of $\alpha_s$ .} .

The "opposite" type fluctuations, which highly enlarge the number
of partons in \Fd compared with the mean value, can be generated
by the following type parton mechanism.  Firstly the valent
partons should emit a secondary parton which is far in transverse
plane - at the distance $ \bp > r_0 Y$, and with the energy $\sim$
the primary particles energy $E$. The probability of such an
emission is $ \sim \exp{(- \mu (r_0 Y) )} $. Then this isolated
parton will give with the probability $\simeq 1 $ an independent
cascade with $\simeq$ the same number of low energy partons as in
the parent cascade. So, the mean number of partons in this state
will approximately double. In the same way emitting two such far
partons at the first stages of cascading (the probability $\sim
\exp{(- 2 \mu (r_0 Y) )}$) we come to the tripling of the number
of low energy partons, etc.

Hence one can simply conclude that the tail of the multiplicity
distribution in Froissart asymptotic collisions  will have the KNO
form of the type
 \bel{mdkno}
 \sigma_N ~\sim~ \exp{(- c~N/<N(Y)>)}~,~~
 c  \sim 1~~~,
 \ee
where $<N(Y)>$ is the mean multiplicity in an inelastic
interaction. It is interesting that this behavior of the
multiplicity distribution is of the same type as it comes from the
simplest summing of pomeron contributions and also of that type as
it is seen in experiments at high accelerator energies.

 \vspace{3mm}
\underline{{\bf \emph{ Fluctuations of the \F disk edges shape }}}
 \vspace{3mm}

The y-development of the parton cascade also contains such
fluctuations which contribute to a distortion of the shape of \Fd
edge - so that it is not a pure circle, but a randomly oscillating
curve. In the physical state of a fast particle this can be
represented as a superposition of \Fd with various shapes.

Such a distortion of the \Fd edge comes from small fluctuations of
the rate of propagation of partons density fronts in the different
directions of the transverse plane
 \footnote{It is clear that the processes creating such
fluctuations are not taken into account in the ``deterministic''
diffusion-type equation (\ref{nleq}) for the mean parton
densities. ~One can effectively include these  fluctuations by
adding to right hand side of Eq.(\ref{nleq}) stochastic noise term
as was done in \cite{peschan}. Such terms are almost inessential
in internal parts of \Fd , but they lead to big fluctuations of
the \Fd edge. }.
 ~For a long time of propagation ~($\sim Y$) of such a front, these
velocity fluctuations accumulate and manifest themselves in a
random shift of the position of the edge of the disk by an amount
~$\sim \sqrt{Y}$.

In the $(2+1_{\bot})D$ QCD the spectrum of shape distortion of \Fd
, as discussed in  Section 2 , has the universal gaussian form
with the average amplitude of the edge fluctuation of order $r_0
\sqrt{Y}$ and here only the value of $r_0$ depends on details of
parton dynamics.~

In the 4D QCD the structure of the \Fd shape fluctuations is
little more complicated but it is of the similar type ~(See
details in Appendix).~~ At large $Y \gg 1$ the average amplitude
of the radial edge oscillation is $\lm(Y) \sim r_0\sqrt{Y} $ as in
the 3D QCD.~

 So  are shape fluctuations of the soft sub-disk. Shape
fluctuations of the more hard saturated sub-disks have  a close
structure. This is due to the fact that hard partons almost do not
move in the transverse plane in the process of their creation in a
cascade from partons of a lower virtuality. As a result, the
fluctuations of the shape of the sub-disk with the virtuality u
are the same as fluctuations of a soft sub-disk with the same mean
radius $\simeq r_0(y - u^2)$. That is, the mean width of
u-sub-disk edge is $\lm(Y,u) \sim r_0\sqrt{Y - u^2}$.

\section*{\bf 4. Collision of two F-disk}
\nfn{4}

In this Section we briefly describe the main properties of the
high energy hadron interaction which on can expect considering it
as a collision of two saturated \Fd  with energies  $E_1 = m
e^{y_1}$ and $E_2= m e^{y_2}$ with the impact parameter b.

The total inelastic cross-section ~$ \sigma_{in} \simeq \pi
(R(E_1) + R(E_2))^2 \simeq \pi r_0^2 (y_1 + y_2)^2 $ is
boost-invariant.
 The elastic amplitude at $Y = y_1 +y_2 \rightarrow \infty$, as on
can expect, is diffractive and purely imaginary
 \bel{elamp}
 A(Y,\bp) ~\simeq~ i ~\tilde{\theta}(r_0 Y - \bp)
  ~~~,~~~~~~~~ \tilde{\theta}() \simeq \theta()~~,~~~~~~
  T(Y,\bp) \simeq ~1 - \tilde{\theta}(r_0 Y - \bp)~~.
 \ee

Since we expect that distributions of the disk radius are gaussian
and diffractive scattering from the different parts of the target
are additive,  the smoothed \F elastic amplitude can be
approximately represented in the form
 \beal{ela4}
A(Y, k_{\bot})~\simeq~  \frac{i \bar{R}(Y)^2}{\lm (Y) \sqrt{\pi}}
        \int~  e^{-\frac{\xp^2}{\lm^2(Y)}}
    ~~\frac{ J_1 \Bigl(k_{\bot}(\bar{R}(Y)+ \xp) \Bigr)}{
      k_{\bot}(\bar{R}(Y)+ \xp)}
     ~d\xp~~,~~~~~~~~~~~~~~ \\
\nn  \bar{R}(Y)~\simeq~ r_1 + r_0 Y ~,~~~
 \lm (Y) ~\simeq~ c r_0 \sqrt{Y} ~,
  ~~~ c ~\simeq~ 1~,~~~r_1 > r_0~.
 \eea
This corresponds to the superposition of contributions to $A$ from
scattering on $\theta$-like profiles distributed around the mean
$\bar{R}(Y)$ and the mean shape of  edge $\lm (Y)$. ~For an almost
black \Fd when the transparency $T \ll 1$ we have $\sigma_{el}
\simeq \sigma_{in} $.

 For the hard scattering when $k_{\bot} \gg 1/R$  the main
mechanism of the elastic scattering is different.  It comes from
components of the wave function of the colliding particles that
contain minimum of partons and no \Fd. This leads to a power
decrease of $d\sigma_{el}/d k_{\bot}^2$ with $k_{\bot}$~.~ Those
cross-sections also contain an additional factor $\sim
\exp{(-\Delta Y)}$ representing the probability that the colliding
particles are in  bare states
 \footnote{Or, more accurately,  $\sim \exp{(-\Delta_1 Y \ln Y)}$
if one takes into account that $\alpha_s$ varies with $u$ }
 that contain no \Fd.

\vspace{5mm}
  \underline{\bf \emph{Inclusive spectra and multiplicity}}
\vspace{3mm}

Inclusive spectrum of secondary particles $\rho_1(y,Y)$ as a
function of rapidity $y$ can be simply estimated as being
proportional to the average transverse area of the two \Fd
intersection in the system where rapidities of the colliding
particles are y and Y-y. This gives the contribution from the
collision of \underline{soft} sub-disks to the created particles
density
 \bel{incl}
 \rho_1(y,Y, u \sim 1) \sim~  \rho_0 \cdot y^2 ~(Y-y)^2 ~/Y^2
  ~~,~~~~~
 \ee
where $\rho_0$ is the parton density in the soft sub-disk. The
corresponding soft particle multiplicity is $\emph{N} = \int
\rho_0 dy \sim Y^3$.~~

The collision of hard sub-disks produces secondary particles with
larger transverse momenta $k_{\bot} = \exp{(u/2)}$. The
corresponding hard inclusive cross-section can again be estimated
to be proportional to the average intersection area of the
sub-disk with the virtuality $u$ with sub-disks of the same or
with a lower virtuality. It must be also proportional to the
parton density in u-sub-disk $\sim \exp{(u)}$. ~Taking all this
into account we receive
 \bel{hins}
\rho_1(y,Y,u) \sim~ \frac{e^u}{Y^2} ~\Big[(y - u^2)^2 (Y-y)^2
~\theta (y - u^2) ~+~
                y^2 (Y-y - u^2)^2 ~\theta (Y-y - u^2) \Big]~.
 \ee
Here the main contribution comes from the collision of saturated
sub-disk with a big virtuality $\sim u$ from one particle with the
soft disk from the other particle.

The full multiplicity of created hard particles with transverse
momenta $k_{\bot} = \exp{(u/2)}$ is $N(Y,u) \sim e^u ~(Y-u^2)^3$.~
So the spectra of these particles grow with u up to $u_{max} \sim
\sqrt{Y}$ and at larger u it rapidly decreases $\sim
\exp{(-(u-u_{max})^2/Y)}$.~ The mean virtuality of these particles
is
 $$
 \bar{u} \sim \int N(Y,u) ~u ~du/\bar{N} \sim Y^{1/2} \sim u_{max}~~,
$$
where $\bar{N} = \int N(Y,u) du \sim Y^{1.5} \exp{\sqrt{Y}}$ is
the full mean particles multiplicity.

Eq. (\ref{incl}), (\ref{hins}) give the spectra of  primary
particles, created when the two \Fd 's move one through another.
The interaction between these particles in the final state can
additionally thermalize the created system with the mean
temperatures $\sim \exp \Big(Y^{1/2} \Big)$.

\vspace{3mm}
 The shape of the multiplicity distribution of the created
particles in the \F limit is defined mostly by the geometrical
reasons (the main dependence comes from the distribution of impact
parameter B), as in the case of collision of heavy nuclei. As it
can be seen from (\ref{fpd}) the mean multiplicity $\bar{N} \sim
\exp{ \big(~\sqrt{Y}~\big)}$ changes only slowly for $0 < \bp <
\bar{R}(Y)/2$ and decreases as ~$\bar{N}(\bp) \sim
\exp{\big(~\sqrt{Y -(2 \bp - \bar{R}(Y))/r_0}~\big)}$ ~for~
$\bar{R}/2 < \bp < \bar{R}(Y)$.

The tail of the multiplicity distribution, as discussed above,
comes from  rare long range fluctuations (the creation of
energetic partons with $x_{\bot} > \bar{R}(Y)$ ) when several ( n
)  mean \Fd 's are created. The probability of these fluctuations
is $\sim \exp{(-n \bar{R}(Y)/r_0)}$. Such fluctuations correspond
to the components of the wave function of a fast particle
containing $ \simeq n
* <N>$ partons, and this leads to the KNO type form (\ref{mdkno})
of the multiplicity distribution.

\vspace{10mm}
  \underline{\bf \emph{ Diffraction generation}}
\vspace{3mm}

Processes of the diffraction generation (~DG~) take place at the
edges of the \Fd 's , when  two  colliding \Fd 's only touch each
other. In this case, the different components of the wave
functions of the colliding particles have different transparencies
and this is the source of the diffraction generation of various
final states. Since the shape fluctuations of saturated \Fd are
big~ $\lambda \sim r_0\sqrt{y}$, it will reflect in the full
diffraction generation cross-section $\sigma_{dg}(Y) \sim
\bar{R}(Y) ~\lambda(Y) \sim Y^{3/2}$.

Depending on the rapidity $y_i $  of colliding particles this
process will look differently.~ In lab frame of one particle $y_1
\simeq 0,~ y_2 \simeq Y$ the probability for the particle 1 to
convert diffractively to another state is $\sim  r_0 \bar{R}(Y)$.
This is because the length of saturated soft edge of \Fd almost
does not fluctuate, and only at the edge of width $\sim r_0$ the
different components of the particle 1 have different
probabilities to interact with the particle 2. ~It corresponds to
the diffraction of the particle 1 in  small mass beams with the
cross-section $\sim Y$~.

At the same time, in frames close to cms when $y_1 \sim y_2$ the
collision of two \Fd with their edges can correspond to
diffraction generation of particles (or beams) at these rapidity
in the multi-regge type configurations (Fig. 4 b,c) with
cross-sections $\sim \lm(Y) \bar{R}(Y) \sim Y^{3/2}$.

The hard  diffraction generation originates from the collision of
edges of hard \Fd sub-disks with $\sigma \sim R(y,u) \simeq r_0 (y
- u^2)$. This process arises  together with the soft production of
particles coming from the collision of the more soft sub-disks.

\vspace{10mm}
\section*{\bf 5. Transparency  of the Froissart disk and\\
     boost-invariance  of cross-sections} \nfn{5}

The condition of boost-invariance (the Lorentz-frame independence)
of various amplitudes and cross-sections calculated in the parton
approach is rather strong. ~Probably it imitates the t-channel
unitarity conditions for the  connected scattering amplitudes and
so can essentially restrict the calculated quantities.~~ Before we
consider the collision of QCD-like \Fd it is interesting at first
to discuss  general restrictions on the parton structure that
follows from boost invariance of cross-sections.~~

 Let us consider the collision of two particles which can be represented
as the partonic clouds that are in a state of a very rare gas. It
is the case normally described by the reggeon diagrams, that, by
their construction, include t-unitarity requirements, so here we
probably should not meet any problems. Let the mean number of
partons in colliding hadrons be $n(y)$, $n(Y-y)$~; and the mean
transverse radii of regions occupied by these partons are~ $R(y)$,
$R(Y-y)$ respectively. Then the total inelastic cross-section can
be represented~as:
 \beal{sts1}
  \sigma_{in} (Y) ~=~ \sigma_0~ n(y)~n(Y-y)~~-~~~~~~~~
  ~~~~~~~~~~~~~~~~~~~~~~~~~\nn \\
  ~~~-~a_2 ~\sigma_0~ n(y) n(Y-y)~\Big( \frac{\sigma_0 ~n(y)}{R^2(y)}
  ~+~ \frac{\sigma_0 ~n(Y-y)}{R^2(Y-y)} \Big) ~+...~~,
 \eea
where $\sigma_0$  is the parton-parton cross-section, ~$a_2 \sim
1$.~ The first term in (\ref{sts1}) corresponds to a collision of
at least one pair of partons. The next terms describe corrections
from screening and multiple collisions.~

For a rare parton gas one can in the first approximation neglect
multiple collisions and screening, that means to leave only the
first term in~(\ref{sts1}). Then, from the requirement of the
independence of ~~$\sigma _{0} n(y) n(Y-y)$~~ on ~$y$~ it follows
the unique solution for ~
  \bel{n0}
 n(y)=n_{0}~e^{ \Delta_0 y }
  \ee
with some real constants $n_{0}$, $\Delta_0$.~ Such a behavior of
$\sigma_{in} (Y)$ corresponds to a regge pole in the complex
angular momentum plane with intercept $\alpha(0) =  \Delta_0$ (and
not a cut or some more complicated regge singularity ). And this
condition follows in the relativistic Regge approach only from the
t-unitarity~ .

If we make the next step and impose the condition of $y$
independence on the sum of two terms in the right hand side of
(\ref{sts1}) and assume that the correction to (\ref{n0}) is
small, we become the corrected expression
  \bel{ny2}
n(y)=n_{0} ~e^{\Delta_0 y}  +~
  a_2 ~n_0^2 ~\frac{\sigma_0}{R^2 (y)} ~e^{2(\Delta_0 y)} ~+...~~~~
 \ee
The second term in (\ref{ny2}) corresponds to the two reggeon
exchange diagram whose structure is almost complectly fixed here
from the boost-invariance. The coefficient $a_2$ depends on the
weight of diffractive amplitudes entering in the two regeon
emission vertexes~
  \footnote{ Moreover, if we consider  the cross-section in
Eq.(\ref{sts1}) with the definite impact parameter $\bp$ then from
the frame independence of the $\sigma_{in} (Y,\bp)$ the form of
transverse parton density $n(y,\bp)$ and of the $R(y)$ is almost
completely fixed in the usual regge pole form  $n(y,\bp) ~\sim~
y^{-1}~ \exp\big( \Delta_0 y -\bp^2/4 c_2 y \big)~,~~R^2(y) \sim y
$.
 }.~
 The possible next terms in (\ref{ny2}), corresponding to higher
regge cuts,  can be found in the same way by the applying the
boost-invariance condition to the larger combinations of screening
terms in expression (\ref{sts1}) for~$\sigma_{in} $.

\vspace{2mm}
 The more suitable example is the high energy collision of fully
black disks (representing fast colliding particles) whose radii
$R(y_i)$ somehow depend on their rapidity $y$ and $Y-y$. In this
case the total inelastic cross-section can be determined from
purely geometrical conditions as
 \bel{st2}
 \sigma_{in} (Y) ~=~  \pi~ \Big( R(y)+R(Y-y) \Big)^2~.
 \ee
From the condition  of the independence of the right hand side of
Eq.(\ref{st2}) on $y$  follows the unique solution for
~$R(y)=r_1\cdot y + r_2$~.~ So, in the case of black disks we
immediately come to the \F type behavior of cross-sections.

But if disks are grey  the picture changes. So, for the case of
constant local disk transparency $T_l  > 0$ we have in lab frame~~
$\sigma_{in} (Y) \simeq \pi R(Y)^2 ~( 1 - T_l)$ and in cms frame
$$
\sigma_{in} (Y,B) \simeq \pi R(Y)^2 ~
 \big( 1 -  c r_0/R(Y)  \big)~~.
$$
This phenomenon is illustrated in Fig.2,~ and it shows that the
grey disk picture of a fast particle is contradictory. The same is
true if local $T$ varies inside disk, as for example in the case
of soft saturated \Fd .
\begin{figure}[h]  \centering
\begin{picture}(160,155)(80,0)            
\includegraphics[scale=.4]{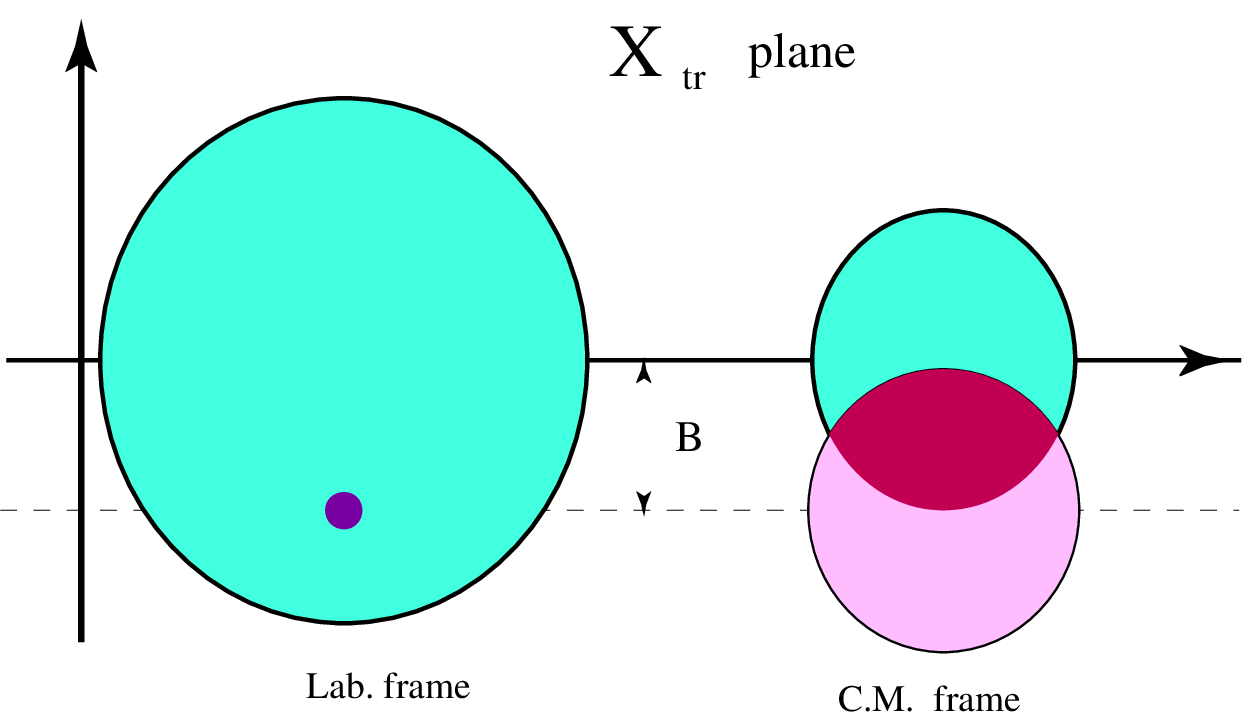}
\end{picture}
\parbox{14cm}{
 {\bf Fig.2}~~ The area of overlapping $S_{12}$
of two colliding Fd, taken at the same impact parameter and total
energy, depends on the longitudinal Lorentz frame. For a {\bf
grey} disks this leads to boost non invariance of the transparency
$T \sim \exp{(-c S_{12}/r_0^2)}$~. } \label{Fig.1}
\end{figure}
On can try to cure this simple model supposing that the fast
particle state is the superposition  $\psi (y) = a(y)
~|disk\rangle + \sqrt{1- a(y)^2} ~|bare \rangle$ of a grey disk
state and of the bare state containing $\sim 1$ partons and
therefore almost does not interact at large $y$. Than one can vary
the amplitude $a(y)$ and try to come to the boost-invariant $
\sigma_{in} (Y)$. It is possible to do  but only for one definite
impact parameter. So going this way one needs to have an infinite
term superposition with a wide distribution of  greyness in disk,
like for a system in the critical point.

\vspace{3mm}
 Considered in the previous Sections the QCD inspired
models correspond to  \Fd's   that are filled by partons so that
they are black in inner parts of \Fd but are always grey near
their edge. Therefore it is interesting to consider what
distribution of the parton density inside such disks can lead to a
consistent boost-invariant picture.

The transparency in a high energy interaction of particles $a$ and
$b$ with rapidity $y_1,~y_2$  can be expressed as
 \bel{gtr}
   T(y_1,y_2,\bp) ~=~ \sum_{i,j} w^{(a)}_i(y_1) ~w^{(b)}_j(y_2)
      ~T_{ij}(y_1,y_2,\bp) ~~~,
 \ee
where we sum over all parton configurations of particles  $a$ and
$b$~. In (\ref{gtr}) $w^{(a)}_i$  and  $w^{(b)}_j$ are the
probabilities of these configurations, and  $T_{ij}$ - the
corresponding transparency in a ~$|i> * ~|j>$  colliding state.
One can expect that for a majority of parton configurations
$|i*j>$ the transparencies  $T_{ij} \sim \exp{(-c N_{ij})}$ are
Poisson-like,~ where $N_{ij}$ is the mean number of parton
collisions in the $~|i> * ~|j>$ scattering.

So, to calculate the full transparency one must sum over all
possible parton collisions in the region where two \Fd -s
intersect.~~ With the exponential precision the transparency can
be expressed through the  saturated parton densities $f_s (y,u,b)$
in \Fd as:
 \bel{trah}
T(y_1,y_2,\bp) ~\sim~  \sum_{i,j} w^{(a)}_i(y_1) ~w^{(b)}_j(y_2)
 \exp \Big( - \tau_{i j} (y_1,y_2, \bp) ~\Big) ~,
 \ee
where the expression
 \bel{gabb}
 \tau_{i j} (y_1,y_2, \bp) ~\sim  \int d^2 \xp \int du_1 du_2 ~
 \sigma (u_1,u_2) \cdot f^{(i)}(y_1, u_1, |{\vec \xp}|)
 \cdot f^{(j)}(y_2, u_2, | {\vec \bp}- {\vec \xp} |)  ~,
 \ee
is proportional to the mean number of the parton scattering when
two \Fd  in the states $|i> * ~|j>$ penetrate one trough another
during their collision at the impact parameter $\bp$ ,~ and
$$
 \sigma (u_1, u_2)  \sim 1/(k_{1\perp} \cdot k_{2\perp})
           \sim  \exp(-(u_1+u_2)/2)
$$
is the cross-section for the parton interaction with virtualities
$u_1$  and $u_2$~ and  $f^{i} , ~f^{(j)}$  - parton densities in
states  $|i> * ~|j>$~.

If in the expression  (\ref{gabb}) for $\tau_{i j}$ we leave only
the terms with small values of the parton virtuality $u_i = u_0
\sim 1$, we come to the case of the of a grey \Fd collision when
 \bel{gabbs}
\tau (y_1,y_2, \bp) ~\sim  \int d^2 \xp ~f_s(y_1, u_0, |{\vec
\xp}|)
 \cdot f_s(y_2, u_0, | {\vec \bp}- {\vec \xp} |)~~.
 \ee
Since inside \Fd the parton density  $f_s(y_1, u_0,\xp)$ is
$\simeq const(\xp)$ the expression (\ref{gabbs}) is evidently not
a boost invariant. It follows from (\ref{gabbs}) that the value
of~ $\tau (Y, 1, \bp) \sim 1$ in the lab frame and $\tau (Y/2,
Y/2, \bp) \sim Y^2$ in the cms system. ~The  phenomenon was
illustrated in the geometrical way in   Fig.2.~ It is of same type
as in the case of QCD in $(2+1_{\bot})D$ , considered in
Section~2.

The expression (\ref{gabbs}) can be  boost invariant only for some
very special forms of $f_s$, for example, for the Gaussian form of
$f_s$ like
 \bel{redf}
  f_s(y, u_0, |{\vec \bp}|) ~\sim~  \frac{1}{y}~e^{\Delta y  -
    \bp^2/y r_0^2}~~,
 \ee
which corresponds to the parton distribution arising in the parton
cascade ( without the parton gluing~ !), The same parton
distribution also corresponds to the amplitudes described by the
regge pole exchange with intercept $\Delta > 0$. ~In fact, such
one expression corresponds again to a black disk of radius $r_0
y\Delta $~ with a thin edge, because here the parton density
changes fast from  small to big values at the distances $\delta
\xp \sim r_0/\sqrt{\Delta}$.

\vspace{8mm}
 \underline{{\bf \emph{5a. ~Central collision of ~\F - disks }}}
 \vspace{5mm}

Consider the central-like collision of two \Fd with a small impact
parameter $\bp$, when
 \bel{dcon}
 | ~\bar{R}(y_1) + \bar{R}(y_2) - \bp ~| ~\gg ~\lm(y_1) +
\lm(y_2)~.
 \ee
The geometry of the collision of two \Fd in various systems and
with the same impact parameter and the total energy looks like
shown in Fig.2~.

Firstly we estimate the contribution to the transparency from the
mean parton configurations.~ In the expression (\ref{gabb}) the
hard partons  give the main contribution to the value of $\tau$,
and using for $f_s$ the expressions (\ref{partde}), we become  :
 \bel{gabbfu}
 \tau (y_1,y_2, \bp) ~\sim~
 \exp \big( ~c \sqrt{y_1 + y_2 - \bp/r_0} ~\big)~,~~~ c\sim 1 ~~,
 \ee
where the main contribution to $\tau$ comes from the collision of
the most hard sub-disks
 \footnote{The expression (\ref{gabbfu}) is estimated with an exponential
precession and it is boost-invariant. The corrections to
(\ref{gabbfu}) are not boost invariant, but this does not change
the main conclusion that in this case the mean density states of
\Fd in 4D case are not essential for the transparency }
  at given $y_i$ and the impact parameter  $\bp$.~

 As follows from (\ref{gabbfu}) and (\ref{trah})  the mean states of
colliding \Fd give no essential contribution to $T$ in all Lorentz
systems, including the laboratory frame.
 At $Y = y_1 + y_2 \gg 1$ this is much less than the contribution
in T that comes from some rare components of the parton wave
function in (\ref{gtr}, \ref{trah}), corresponding to the
particles  states which have smaller transverse sizes so that they
interact weakly at the same $\bp$.~~ These are basically the
states containing no \Fd, or \Fd of smaller radius. For a
collision at the impact parameter $\bp$ one needs the states in
(\ref{gtr}) with ~$(R(E_1) + R(E_2) < \bp$ in which the \F disks
do not touch each other. It is simple to estimate their
probability $w_i$, because these states are basically created by
fluctuations in which the primary \underline{soft} partons are not
emitted at the rapidity intervals $\delta y_1, \delta y_2$,~ close
to the colliding particles valent zones, so that $ (Y -\delta y_1
- \delta y_2) r_0 < ~\bp$. This condition is boost-invariant, and
the corresponding transparency is
 \bel{trnof}
 T \sim  w^{(a)} \cdot  w^{(b)} \sim  e^{- \delta y_1}
\cdot  e^{- \delta y_2} \sim
 \exp{( - c_1~(Y - \bp/r_0))}~~,~~~  c_1 \sim \alpha_s~,
 \ee
and has the same property.

Therefore, ~for the collision of the \Fd with such a ~$\bp$~ one
probably can avoid problems  with boost-invariance of T~.

 \vspace{10mm}
\underline{{\bf \emph{5b. Two \Fd disks collision close to there
edges }}}
 \vspace{5mm}

\nin Note that  expressions (\ref{trah}, \ref{gabbfu}) for the
transparency  $T(y_1,y_2,\bp)$ become unapplicable for large
impact parameters where
 \bel{strip}
 |~\bar{R}(Y) - \bp~| \sim \lm  \sim r_0 Y^{1/2}
 \ee
that is in the stripe where the \Fd edge fluctuates. Here
different mechanisms for the transparency T can operate.

For such a $\bp$ in the { \bf lab frame} of one particle  we have
$T_{lab} \sim 1$, and it approximately do not change with growth
of Y~.~This is because there are such curved configurations of \Fd
edge of another fast particle (and which appears with the
probability $\sim 1$) that particles do not touch each other
(Fig.3a).

\vspace{10mm}
 \begin{figure}[h]  \centering
\begin{picture}(160,175)(80,20)
\includegraphics[scale=.5]{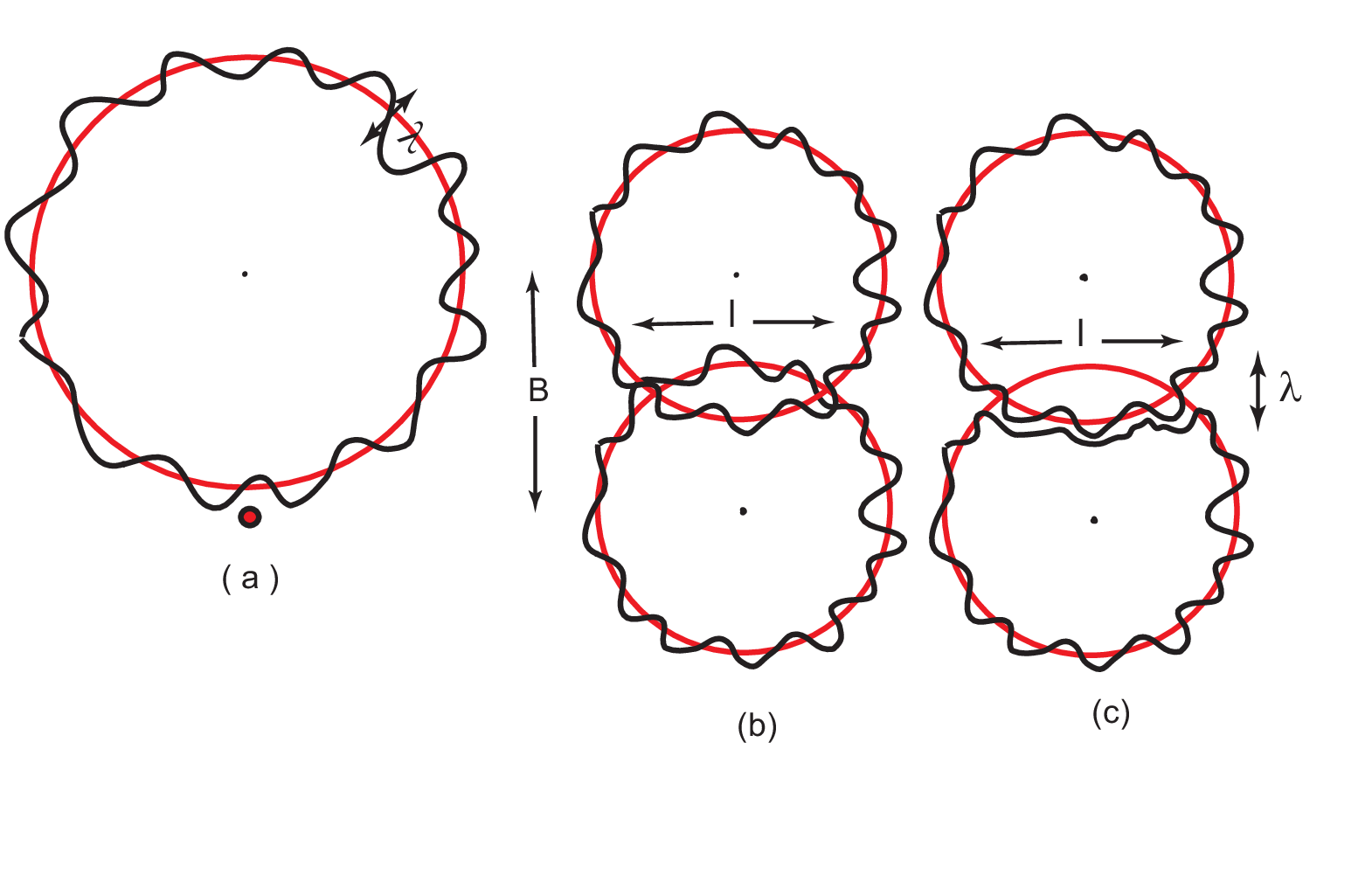}
\end{picture}
\parbox{14cm}{  {\bf Fig.3}~~ Collision of two \Fd near their
edges in lab. (a) and cms. (b and c) frames at the same impact
parameter and total energy. Red line round corresponds to disks
with average $\bar{R}(y)$, and black oscillating lines to real
random configurations in the mean event. The  Fig.c shows a rare
configuration of one of the \Fd where disks do not interact.}
\label{Fig.1}
 \end{figure}

But consider collision at the same impact parameter $\bp$ and in
the {\bf frames close to cms} when the two \Fd intersect in the
mean configurations (Fig.3b). The average width of this
intersection region is $\sim \bar{\lambda}(Y) \sim Y^{1/2}$,~ and
the mean tangential length $l \sim \sqrt{ \bar{R}(Y)\lambda} \sim
Y^{3/4}$.~ This leads to the transparency
 \bel{tosc1}
 T_{cms} \sim \exp{(- c~l \lm/r_0^2 )} \sim
 \exp \Big(-c~ \lm^{3/2}(Y) ~R^{1/2}(Y)/r_0^2  \Big) \sim~
 \exp{(- c~Y^{5/4})}
 ~,~~~~c
 \sim 1
 \ee
in the mean \Fd configurations.

Next, one must estimate the probability $w$ of a special big
fluctuation of part of the edge of \Fd of length $\sim l$, which
is such that here the edge line shifts to smaller radius on the
value $\sim \lambda(Y)$. In such configurations $T \sim 1$ also in
cms at the same impact parameter (Fig.3.c). This probability can
be estimated (from above !) in such a way. In the  mean \Fd
configurations the radial position of the \Fd edge randomly
fluctuates around the average $\bar{R}(Y)$ when we move along the
edge. It moves in both sides,~ $R > \bar{R}(Y)$ and $R <
\bar{R}(Y)$~, shifting on the mean radial distance $\sim \lm(Y)$.
During this ``motion'' it crosses the line $R = \bar{R}(Y)$,  and
the average number of such ``long wave'' intersection at the
distance $l$ is $\sim \sqrt{l/\lm(Y)}$
 \footnote{In fact, the mean number of all such intersections is
much larger $\sim \sqrt{l/r_0}$. ~Note that for $l \sim R(Y) \sim
r_0 Y$ it gives $w \sim \exp{(-Y^{1/2})}$ . It coincides with the
estimate (\ref{flucy}) of the probability that the radius of the
\Fd is less   by the value of $\Delta \bar{R} \sim \lambda(Y) \sim
\sqrt{r_0 \bar{R}}$   than the average $\bar{R}(Y)$. }.
 ~~Since such crosses are independent their number is Poisson
distributed. So, the probability that at the length $\sim l$ there
are no intersections is $w \sim \exp{(- \sqrt{l/\lm} )}$~~
 \footnote{~Another estimation of the probability of such large
variation of \Fd edge is given by the expression  (\ref{defor}).
Substituting there the dependence of $\lm $ and $l$ from Y we
become $w \sim \exp{(-Y^{1/4})}$. }.
 ~~One can take this $w$ as an estimation of the probability of Fock
component of one \Fd where the part of the parton disk is shifted
inside at the length l , so that two \Fd do not collide also in
cms.~
 It gives for
 \bel{tosc2}
 T_{cms}  ~<~ w \sim \exp{(- \sqrt{l/\lm} )  }
 \sim \exp{(- Y^{1/8} )}~.
 \ee
This contribution to the transparency $T_{cms}$ is bigger than
(\ref{tosc1}) that comes from the mean parton configurations of
\Fd, but it  also {\bf decreases with Y }, and so it cannot
coincide with the transparency in the lab frame at $\bp \simeq
\bar{R}(Y) \pm \lm (Y)$, which is const(Y).

Therefore, here, as in $(2+1_{\bot})$ case, we  have a
contradiction between the naively expected  4D QCD  picture of \Fd
and the boost-invariance (t-unitarity) of some amplitudes.

\vspace{10mm}

\section*{\bf 6. The Froissart type behavior in the  regge approach \\
and  diffractive processes}
\nfn{6}

 \vspace{5mm}
At the first sight, it is natural to expect that the calculations
of various cross-sections in the parton approach should, in
principal, give an answer that is the same as that one can become
using the reggeon approach. For example, so is the transparency
$T(Y, \bp) = 1 -\sigma_{tot}(Y,\bp) + \sigma_{el}(Y,\bp)$, which
we estimated in the parton approach for the case of the collision
of two \Fd, and have found that it can be not boost-invariant for
some values of $\bp,Y$.~ But if we calculate such quantities as
$\sigma_{tot}(Y,\bp) = Im A(Y,\bp)$ and $\sigma_{el}(Y,\bp) =
|A(Y,\bp)|^2$~ ~using reggeon diagrams  for elastic amplitude
$A(Y,\bp)$~, ~we will find that they are by construction
boost-invariant, because the the particular system in which
particles collide does not enter the Lorentz-invariant
calculations of the $A(Y,\bp)$ ~and therefore of the $T(Y,\bp)$.

But possibly there is no contradiction here.~~
 When the pomeron density  is small (in this case $T(Y,\bp)$ can be
large) the parton calculation gives the same boost invariant
answer as in the regge case. This is the consequence of the
Gaussian type form of the distribution (\ref{redf}) of the
transverse parton density in \Fd.

The Regge approach can be consistently applied only when the
reggeon (pomeron) density in the transverse plane is small.
Otherwise, the mean energies at pomeron lines entering dominant
reggeon diagrams are small.  This is what takes place near the
saturation point.  And here, strictly speaking, we are out of the
region of applicability of regge approach. But for parton approach
the big QCD-parton density at $Y \gg 1$ can be even in advantage,
because here we can apply the quasiclassical approximations for
calculations of various cross-sections.

Despite the fact that the large quantities like
$\sigma_{tot}(Y,b)$ have in both approaches roughly the same
properties the small corrections to them (like that contribute to
T) can differ.

But such a inconsistency suggest that if we, nevertheless, apply
the regge approach with the \F asymptotical behavior then we can
meet some direct violations of unitarity. This can be expected
firstly in processes where the interaction of grey parts of \Fd is
essential - it is in some diffractive processes.

 \vspace{3mm}
To estimate such  cross-sections we need the pimary regge
amplitudes that correspond to the collision of \Fd-s.~ The
simplest way how the \F behavior appears in the regge approach
comes from the eiconal summation of contribution of supercritical
pomerons which are the natural objects in QCD.
 The supercritical pomeron (with intercept $\Delta_P > 0$ and slope
$\alpha'_P$)  when we write its contribution to an elastic
scattering amplitude in the impact parameter representation as
 \bel{pom1}
v(Y, b ) = \frac{ig_a g_b}{\alpha^{\prime}_P Y_1 } \cdot \exp
\big(~ \Delta_P Y_1  ~-~  \bp^2 / 4  \alpha'_P Y_1 ~\big) ~,
~~~~~Y_1 = Y + i \pi/2~~,
 \ee
after the eiconal-like unitarisation of the S-matrix, leads to the
expressions
 \beal{eic1}
  S(Y, \bp ) ~=~ e^{i v} ,~~~~~~~
  A(Y, \bp ) ~=~  i ( 1  - e^{i v}   ) ~~,~~~~
 \sigma_{in}(Y, \bp) ~=~   1  - e^{- 2 Im~v}  ~,
 \eea
which have a very distinctive property
 \footnote{The explicit eiconal dependence $ S[v]=e^{iv}$ of
S-matrix on $v$ is, by itself, not essential to reach the \F
behavior, and instead of (\ref{eic1}) one can in the same way work
with generalized eiconal series
 $ A(Y, \bp ) ~=~  i \sum_{n=1}^{\infty} c_n (-v)^n /n!~, $
where $c_n$~ are nearly arbitrary positive coefficients
representing the contribution of diffraction generation beams.
 }.
 For a  large $Y$ from (\ref{pom1} - \ref{eic1}) it follows
approximately that
 \bel{teta}
 \sigma_{in}(Y, \bp)
 ~\simeq~  \theta \Big(~R(Y) - \bp~\Big) ~~,
 \ee
so that expressions (\ref{pom1}-\ref{eic1}) correspond to an
almost black disk with a radius
 \bel{rteta}
 R(Y) \simeq  \cdot 2\sqrt{\Delta_P ~\alpha'_P} ~\Big( Y -
 \frac{ln Y}{2 \Delta_P} \Big)  ~~~,~~~~ Y \gg 1 ~~~,
  \ee
whose grey edge is spread on the value $\lambda(Y)  \simeq \sqrt{
\alpha'_P/\Delta_P}$.~ It expands almost linearly with $Y$, and
thus leads to the \F type behavior of cross-sections
~$\sigma_{tot} = 2~\sigma_{in} \simeq \pi~R^2 (Y) \sim Y^2$. ~
There is no parton density saturation in this case - only the
parton screening in the collision processes. The central inclusive
cross section and the mean particles multiplicity grow as ~$\exp
(~ \Delta_P Y ~) $~. ~~The edge parts of \Fd with nonzero
transparency (grey) are responsible for diffractive processes.
 Since in (\ref{pom1}) the transverse parton distribution in $b$
has very specific `` Gaussian'' shape~ $\sim |v(Y, b )|$ ~with the
thin grey \Fd edge the corresponding  transparency is
boost-invariant.~ But for the amplitudes corresponding to more
grey \Fd than (\ref{pom1})  we can meet troubles with unitarity,
especially for a processes of central diffractive production in
the multipomeron configurations.

To discuss if we have a consistent description of physics we
calculate below cross-sections for processes of the central
diffraction~ Fig.4~~ using the eiconal expression (\ref{pom1} -
\ref{eic1}) for $A(Y,b)$ or its saturated generalization as a
primary regge amplitude (Pomeron $\rightarrow$ Froissaron) in
reggeon diagrams.

\vspace{3mm}
 Let us consider a number of cases.

 \vspace{2mm} \nin {\bf *} ~~
\underline{ For a purely grey \Fd } , when ~$\sigma_{in}(Y, \bp) <
1$~ for all $|\vec{\bp}|$~, the simple estimate shows that the
total cross-section of one particle central diffraction ~(Fig.4b)~
is
 $$
\sigma_{dif}^{(1)}(Y) \sim Y^2 ~Y^2 ~Y = Y^5 ,
 $$
which exceeds the Froissard bound at $Y \rightarrow\infty$. Here
for the case of grey \Fd all impact parameters $b$ contribute and
this gives the first factor $Y^2$. Two additional powers of $Y$
come from the integration over the transverse coordinates of the
created particle~ and one more $Y$ from the integration over their
rapidity. Since the \Fd are grey this contribution is not
cancelled by the overlapping diagram in Fig.4b .
\vspace{10mm}
\begin{figure}[h]  \centering
\begin{picture}(190,225)(80,-10)
\includegraphics[scale=.6]{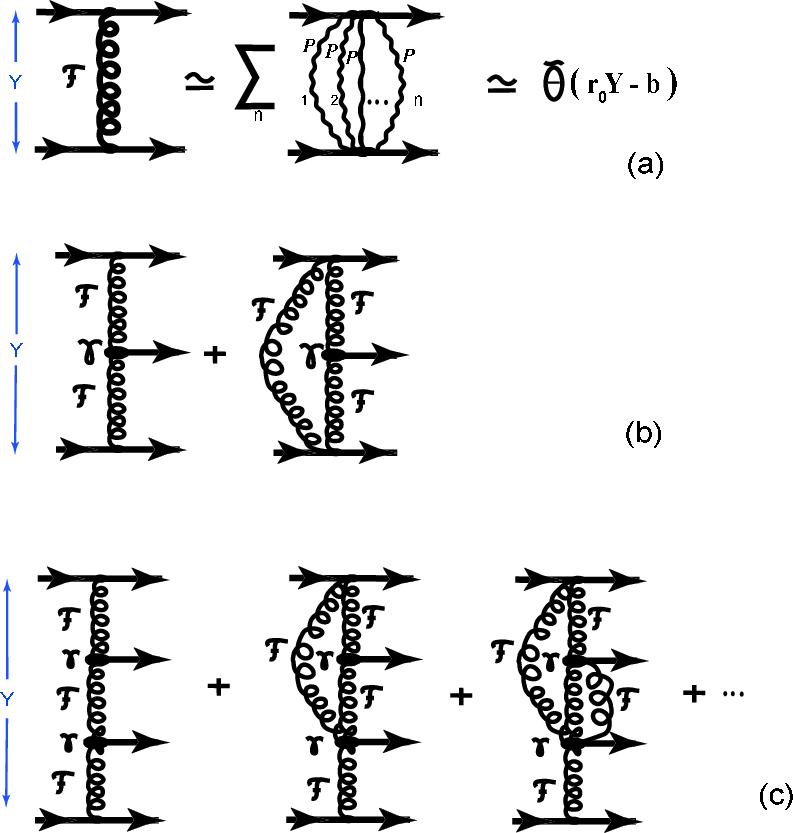}
\end{picture}
\parbox{14cm}{  {\bf Fig.4}~~
(a) - The simplest Froissart amplitude \F can be constructed by
the eikonal  summing of multiple supercritical pomeron
exchanges.~;~ (b),~(c) - Reggeon diagrams  that contribute to a
central diffraction with the \F exchange.}
 \label{Fig.4}
\end{figure}

\nin The integrated cross section for the $n$ particle multi-regge
central diffraction grows here even faster
 \bel{sign1}
\sigma_{dif}^{(n)}(Y) ~\sim~ Y^{3n+2} .
 \ee
This contradiction with unitarity for the grey \Fd  can be avoided
if, ~for example, the vertexes for the central diffractive
particle emission are
 \bel{dver}
\gamma ( k_{1 \bot }, k_{ 2 \bot } ) ~\sim  \vec{k}_{1 \bot }
~\vec{k}_{2 \bot }
 \ee
 at small transverse momenta $k_{i \bot}$ on the \F lines entering the
$\gamma$-vertex. In this case we have $\sigma_{dif}^{(1)}(Y) \sim
Y$.  But such a requirement needs the fine tuning of fundamental
parameters of the theory and is by itself not fulfilled. In fact,
to regulate the behavior of all $\sigma_{dif}^{(n)}(Y)$ one must
have an infinite number of conditions on different vertexes of the
theory, and this can only be result of some special symmetry of
reggeon theory. ~This growth of cross-sections for the \F behavior
of amplitudes reminds the growth of the multiregge diffractive
processes for the asymptotically constant cross-sections
\cite{vkmptm}.

 \vspace{2mm} \nin {\bf *} ~~
\underline{ For the black \Fd  with a grey edge of large width
$\lambda(Y) \gg ln Y$ } the main contribution to the cross-section
of one particle (with rapidity $y$) central diffractive production
(Fig 4 b) comes from configurations when two \Fd with $R(y)$ and
$R(Y-y)$ touch with their grey edges, so that the essential impact
parameters are $\bp \simeq R(Y) \pm \lambda(y_i)$. This gives
 \bel{sign1}
  \sigma_{dif}^{(1)}(Y) ~\sim~
    [~ \lambda(Y)~ (Y \lambda(Y))^{1/2} ~] ~ R(Y)~Y~~  ~\sim~
    ~Y  ( \lambda Y)^{3/2}
 \ee
where the first factor [ $\lambda ~(Y \lambda)^{1/2}$~] gives the
mean area of the transverse region where two \Fd intersect with
their grey parts - the diffractive particle can be created from
every point of this region. ~The additional $R(Y)$ corresponds to
the mean length of the grey edge of \Fd and the last factor $Y$
comes from integration over rapidity $y$ of the created particles.
The inner (black) parts of \Fd give no contribution to $
\sigma_{dif}^{(1)}$ , because there the contributions of two
diagrams  (Fig 4 b) cancel one another. ~~The expression
(\ref{sign1}) for $\sigma_{dif}^{(1)}(Y)$ can be simply
generalized to the case of $n$ central  diffractive (multiregge)
particles production cross-section
 \bel{signd}
  \sigma_{dif}^{(n)}(Y) ~\sim~ [~ \lambda (Y \lambda)^{1/2}~]^n
    ~Y^n ~R(Y) ~\sim~  Y (Y \lambda)^{3n/2}
 \ee
We see from here that any model of \Fd with a ``large grey disk
edge~ (~$\lambda(Y) \gg ln Y$~) leads to a contradiction with
unitarity.

\vspace{2mm} \nin {\bf *} ~~
\underline{ The saturated \Fd in QCD } is mostly black, but it has
the grey edge of the width $\lambda(Y) \sim \sqrt{Y}$~.
 ~The black parts of \Fd  give no contribution to
$\sigma_{dif}^{(n)}$ - there is the screening type cancellation
between different reggeon diagrams.~~
 Thus, from (\ref{signd}) we become for the case of saturated \Fd
 \bel{sign3}
\sigma_{dif}^{(n)}(Y) ~\sim~ c_n ~Y~ (~Y~)^{9n/4} ~
 \ee
where $c_n \sim \gamma^n/(n+1)!$ ~do not depend on $Y$.
 So, here we also have a contradiction with unitarity of the same
type as in the case of grey \Fd  and need to impose some
conditions, may be of type (\ref{dver}). ~Only then we become an
acceptable behavior for all
 \bel{sindif}
\sigma_{dif}^{(n)}(Y) \sim  \hat{c}_n Y \lambda~.
 \ee
\vspace{2mm} \nin {\bf *} ~~
 For the \underline{ black \Fd  corresponding to the eiconal regge
amplitudes (\ref{pom1} -\ref{eic1})} - or to \Fd in QCD without
saturation situation is very interesting ~\cite{kmr}.~ Here we
have the width of the \Fd edge non grooving with $Y$ . And, as
follows from (\ref{teta}), due to the condition $R(Y)-2 R(y/2)
\sim  ln Y$ , the contribution from a configuration when two \Fd
touch with their grey edges are also cancelled with the enveloping
diagram in (Fig 4 b).  So, for estimation  of $\sigma_{dif}^{(1)}$
in this case we can use the expression (\ref{sign1})  with
$\lambda \sim 1$ and with the additional factor
 \bel{fact2}
  1 -  \sigma_{in} \Big( Y, ~R(Y - \frac{ln Y}{2 \Delta_P}) \Big)
 ~\sim~
\Big| ~v\Big(    Y, ~R(Y - \frac{ln Y}{2 \Delta_P}) \Big) ~\Big|
 ~\sim~  Y^{-3/2}
 \ee
and in the same way the additional  factor  $Y^{-3n/2}$  ~for
$\sigma_{dif}^{(n)}$ in (\ref{signd}). ~~As a result, in this case
we come to the unitary expressions (\ref{sindif}) for all
$\sigma_{dif}^{(n)}$ not imposing any additional condition on
vertexes.

\vspace{10mm}
\section*{\bf 7. Conclusion}
\nfn{7}

In the first part of this article ~({\bf Sections 3 - 4})~ we
reviewed briefly the main properties of processes in the case of
the Froissart~(\F) asymptotic behavior, which one can expect in
QCD~ and qualitatively described the parton structure of the
Froissart disk (\Fd) corresponding to a asymptotically fast
hadron.

\vspace{3mm}
  The main aim of this article was to consider if it is
possible to have a boost-invariant picture of the collision of two
\Fd in QCD , ~so to avoid problems with the boost-invariance of a
cross-section of high energy interaction, which necessarily
appears in the case of grey \Fd .
 For this purpose we calculated the mutual transparency
 $$
 T(p_1,p_2, \bp) = 1 - \sigma_{in}(s,\bp) = |S(s,\bp)|^2
 $$
in the process of collision of two \Fd with momenta $p_1$ and
$p_2$ at various longitudinal  systems (ranging from the
lab-system to the cms system) and at the same impact parameter
$\bp$ and total invariant energy $s=(p_1 + p_2)^2$. This quantity
$T$ is equal to the probability that two \Fd move one through
another without interaction - that is both \Fd remain in the same
state as before the collision. The evident requirement is that
such T should be boost invariant - i.e. depends only on s. In
parton approach this condition is not trivial, because particles
state change by a boost transformation. This property of T and
other cross-sections, when calculated in parton approach, probably
reflects the t-channel unitarity conditions of corresponding
scattering amplitudes.

We found that $T$ is not boost invariant for collisions of
saturated \Fd at such impact parameters when two \Fd collide with
their grey parts - it is close to their edges. At these impact
parameters various processes of diffraction generation also take
place. We found that the similar problems with unitarity appear
for these processes in the regge approach in the \F regime.

\vspace{2mm}
 We end with few remarks.

\vspace{2mm} \nin {\bf *} ~~
The \F behavior in the  (2+1)D case which we considered in
Section 2,  as a simple example,  encounters the same problems. It
corresponds to the saturated  grey \Fd and so leads to a  boost
non-invariant S-matrix. This again can signal  that such a
behavior breaks the t-channel unitarity. Possibly, the problems
can be avoided if the parton system is in the critical state, or
in cascade partons only split (and not join) and, therefore, their
density does not saturate. But even if this is true, it still
seems rather strange and can be probably realized only if the
longitudinal distribution in parton cloud is completely different,
such that its mean size grows with energy ~\cite{longp}.

\vspace{2mm} \nin {\bf *} ~~
In the 4D QCD the \Fd becomes almost black in the central part,
but the edges of \Fd are grey and exhibit large fluctuations ``
from event to event''. As a result, for such impact parameters
when two \Fd during their collision  impact only by their wide
edges the value of $\sigma_{in}(s,\bp)$ calculated in the parton
model is not boost invariant and  essentially differs in lab and
cms frames.

To avoid this contradiction the \Fd edge must be thin and
specially arranged
 \footnote{ For example, if the black \Fd profile is such as
results from the summation of the contributions eikonal diagrams
with supercritical pomerons. },
  or oppositely include all the \Fd disks. But the standard QCD
picture of \Fd  is usually considered differently.

 The other possibility is that some very specific interference
between different channels of interaction take place, so that the
resulting transparency T is boost invariant. One needs for this
infinite number of relations between multiparton interaction
amplitudes which are not seen in the quasiclassical picture. The
same situation one sees in the regge approach to calculations of
diffraction generation cross-sections ( Section 6), where one must
impose the infinite number of constraints on pomeron vertexes to
preserve unitarity.

\vspace{2mm} \nin {\bf *} ~~
All this shows that the high energy parton structure of a fast
hadron in QCD must have some very special properties,  otherwise
the usually expected Froissart type behavior should not take place
asymptotically. ~In principle, there is a number of possibilities,
such as mentioned at the end of  Section 2 for the case of
$(2+1_{\bot})$ dimensions, but they all look rather artificial.

\vspace{2mm}
 If for a while forget about the hard component of \Fd, then the
case of the critical Froissaron with grey \Fd seams more promising
(some details are in \cite{Kan1} ).~ It is possible that such a
behavior can be found in the regge approach as a limiting case of
the critical pomeron \cite{ABMPT}.  In this case we can have a
large fluctuation of parton density  of all sizes up to
$\bar{R}(Y)$. ~Moreover, because the soft components of \Fd evolve
independently from the hard ones, and the distribution of hard
component fluctuation can ``repeat'' the distribution of the soft
one. If this really takes place, one can have also the cms
transparency $T \sim 1$.~ But the critical \Fd can meet with fine
tuning of various pomeron parameters in the regge approach.~~
Maybe, some effects in the nonperturbative QCD can make such a
model of \Fd more natural. It would be interesting if there were
also traces of such fluctuations in the data already at
accelerator energies.

\vspace{8mm}
\nin {\bf ACKNOWLEDGMENTS} \\
\nin I thank K.G.Boreskov for interesting discussions and
I.N.Kancheli for help and advice.

\vspace{5mm}
\section*{{\bf  Appendix} \\
  {\textit{Structure of the \F disk edge }}}
\nfn{A}

The process of filling the \F disk  with partons with increasing
of fast particles energy $E = \exp{(y)}$  can be represented as
coming from the development of a parton cascade and resulting from
the emission of the additional less energetic partons by more
energetic ones.

In the process of the parton cascading new partons randomly move
in the transverse plane. For a big number ~$(\sim y)$~ of
cascading steps they fill the transverse area  which is
approximately round, but with small fluctuations of \Fd radius
$R(y,\varphi)$ in  different transverse directions $\varphi$ .
~Here we discuss only the fluctuations in the soft part of the
saturated \Fd at large y , and also suppose that the amplitude of
these fluctuations $\lambda(y,x)$ is small compared to the mean
\Fd radius $\bar{R}(y) =r_0 y$.   The random function
$\lambda(y,x)$ depends on $y$ and the transverse coordinate $x =
\bar{R}(y) * \varphi ,~~ 0< \varphi < 2\pi $ which varies in the
transverse plane along the edge line of the mean \Fd . This
variable x is more appropriate, and so we represent the $R(y,x) =
\bar{R}(y) + \lambda(y,x)$. In fact the \Fd edge, if considered as
a continuous function randomly growing with y , can have  the
complicated fractal structure \cite{surgrow}, but we use here the
smooth approximation.

We need  the weight $W[\lambda(y,x)]$ of the realization of some
definite configuration $\lambda(y,x)$, so that  various mean
quantities depending on  $\lambda$  can be represented by the
averaging
 \bel{bordav}
 \int D \lambda(y,x)~
  W[\lambda(y,x)]~\big(~ \lambda(y,x_1)\cdot\lambda(y,x_2)\cdot ~...\big)
 \ee
of corresponding functions of $\lambda(y,x)$.~~~

For a  smooth  edge $R(y,x)$ and due to absence of long range
interactions in the dense parton medium, the only existing
parameter at the scale $x \gg r_0$ is the length of the edge. In
addition, the far regions of the \Fd edge fluctuate independently.
Therefore, one can conclude  that the amplitude ~$W[R(y,x)]$
depends on $\Gamma [R] - 2\pi \bar{R}(y)$ in the exponential form
 \bel{borac}
 W \sim \exp (-\frac{\beta}{r_1}(\Gamma [R] - 2\pi \bar{R}) ) ~,
  ~~~\beta \sim 1~~~,
 \ee
where $\Gamma [R]$ is the length of the \Fd edge  and $r_1$ is the
average radial distance where the parton density in \Fd passes
from the saturated phase to an unsaturated one ( $r_1 \sim r_0
/\alpha_s$ in the perturbative QCD~ ;~ in the future, for
simplicity, we put $r_1 \simeq r_0$).

At a small and smooth $|\lambda(y,x) | \ll  \bar{R}$ the length of
the edge is
 \bel{gapr}
 \Gamma [R] ~=~
 \int_0^{ L(y)}  dx~ \sqrt{1 + (\lambda '(y,x)_x)^2} ~\simeq~
 L(y) + 1/2 \int_0^{ L(y)}  dx~ (\lambda '(y,x)_x)^2 ~~,
 \ee
where $L = 2 \pi ~\bar{R}(y)$.~ And, therefore,
$$
 W[\lambda] \sim \exp{\Big( - \frac{\beta}{2 r_0}
 \int_0^{L(y)} dx ~\Big( \frac{\d \lambda}{\d x}  \Big)^2
~\Big)}~~.
$$
One can use this expression to estimate the probability $w$ of the
large deepening of the \Fd edge with length $l$ and depth
$\lambda$,~ and we get as a result
$$
  w \sim \exp{(-\frac{\beta}{r_0} \delta \Gamma  ) }~~,
$$
where $\delta \Gamma  = \Gamma [R] - L(y)$ is  variation of length
of the deformed edge. For a long and smooth deformation of edge
from (\ref{gapr}) when $\lambda(x) \ll l$ we have $\delta \Gamma
\sim <\lambda^2>l $,~ and this gives
 \bel{defor}
     w \sim \exp{(-c<\lambda^2>/l r_0)} ~,~~~~~~c \sim 1~.
 \ee
 It follows hence that the average variation of \F disk radius is
  \bel{defor1}
 (\delta \bar{R} )^2  ~\sim~ <\lambda^2> ~\sim~  L r_0  ~\sim yr_0^2
 \ee
The other approach to calculations of (\ref{bordav}) is to
decompose the shape of the edge line in harmonics
$$
 \lambda(y,x) ~=~ \sum_n ~a_n ~e^{i \frac{2\pi n x}{L}} ~~.
$$
Then we have for
$$
\Gamma [R]  = L + \frac{\pi^2}{L} \sum_n n^2 a_n^2~~,
$$
and, therefore, for the averaging over the edge fluctuation we can
use the normalized  measure
 \bel{daw}
 \int D \lambda(y,x)W[\lambda]~(...)  \equiv \int Da~(...) =
 N_1! ~(\frac{\beta\pi^2}{r_0  L} )^{N/2}
 \int_{-\infty}^{\infty} \prod_{n=1}^{N_1} da_n
 \exp {(- \frac{\beta \pi^2}{r_0 L} ) \sum_n n^2 a_n^2)}(...)~,
 \ee
 where we cut the series of harmonics at large $N_1 > L(y) \gg 1$.

Using (\ref{daw})  we can  calculate various quantities that
characterize the \Fd edge, for example, the mean width of the edge
 \beal{delR}
 <(R(y)-\bar{R})^2> ~=~
\frac{1}{L(y)}~< \int_0^{ L(y)} dx |\lambda(y,x)|^2 >  ~=~
 \frac{1}{2}~\int Da ~ \sum_{n=1}^{N_1}  a_n^2  ~=~  \nn \\
 =~ \frac{\beta}{4\pi^{3/4}} ~r_0 L(y)
 ~\sum_{n=1}^{N_1} \frac{1}{n^2}    ~\simeq~
  \frac{\zeta(2)\beta}{4\pi^{3/4}} ~( r_0 L(y))~
  \sim~ r_0^2 ~y  ~~.~~~~~~~~~~~~~~~~~~
 \eea
This shows the same behavior of the width of the \Fd edge as in
(\ref{defor1}) and in the $(2+1_{\bot})D$ case.

The instructive quantity that shows the structure of the \Fd edge
is the correlator \\ $ G_y(x) = <| R(y,x)-R(y,0)|>$ , and
especially its dependence on $x$. From the equations above we have
:
$$
 G_y(x) ~=~
 \int Da ~ \sum_n ~\sin(\frac{2\pi n x}{L})~|a_n| ~\simeq~
 \pi^{-3/2} \sqrt{\frac{r_0 L}{\beta}}
  \sum_{n}\frac{1}{n} \sin(\frac{2\pi n x}{L}) ~\simeq~
$$  $$
 ~\simeq~ \pi^{-3/2} \sqrt{\frac{r_0 L}{\beta}}
   \sin(\frac{2\pi x}{L})
  ~\simeq~
 \frac{2~x}{\sqrt{\pi \beta}}  ~\sqrt{ \frac{r_0}{L} }
$$
At $x \sim L$ we have $ G_y(\sim L) \sim \sqrt{L r_0}$ , which
agrees with (\ref{delR}),~  and shows that the growth of the \Fd
edges width comes from the long range harmonics.

The different approach to the description of \Fd edge fluctuations
is to start from the stochastic parton cascade evolution of
$W[\lambda]$ in rapidity, or, even simplier, from the evolution of
the local ~~(in x) radius of the \Fd edge line in rapidity
 \bel{dry}
 \frac{\d R(y,x)}{\d y} ~=~ \hat{r}  ~-~ c ~r_0^2 ~\rho (y,x) ~,
  ~~~~~c \sim 1~~~,
 \ee
where the  stochastic function $\hat{r}(y,x) > 0$ , with
~$<\hat{r}^2> \simeq r_0^2$~,~ represents the random motion of the
edge due to a parton splitting near the edge of \Fd. The second
term in (\ref{dry}) is  proportional to $\rho (y,x) ~\sim~ \d^2
R(y,x)/\d^2 x$ ~-~ the local curvature of the \Fd edge line. This
term takes into account that the rate of a parton creation close
to the edge is proportional to the number of neighboring partons.
For the convex part of the \Fd edge $\rho
> 0$ and for the concave part  $\rho < 0$~.~~
Averaging Eq. (\ref{dry}) over $x$ we become the equation for the
evolution with y of the mean \Fd radius
 \bel{drym}
 \frac{\d \bar{R}(y)}{\d y} ~=~ r_0 - ~c~ \frac{r_0^2}{\bar{R}(y)}
 \ee
Its solution
 \bel{rlog2}
 \bar{R}(y) ~\simeq~ r_0 ( y - c~ \ln y ) ~~,~~~~ y ~\gg~1~~
 \ee
contains the~  $-\ln y $ corrections, coming from the positive
curvature of the average \Fd edge, the same as in Eq.
(\ref{teta}), that corresponds to the supercritical pomeron pole.

\nin One can also use Eq.(\ref{dry}) to estimate the growth with
$y$ of
$$
\mathcal{G}(y) ~=~ \langle~ (R(y,x) - \bar{R}(y))^2 ~\rangle_{
\hat{r}} ~~~,
$$
which gives  the dependence of the mean width of a \Fd edge on
$y$. Using (\ref{dry} , \ref{drym}) we become
$$
 \frac{\d }{\d y} \mathcal{G}(y)
 ~\simeq~  2 \langle~  \hat{r}(y,x) R(y,x)  ~\rangle_{ \hat{r}} - r_0
 \bar{R}(y) ~=~ 2 \langle~ \hat{r}~\lambda  ~\rangle_{ \hat{r}}
 ~=~ c_3 ~r_0^2  ~~~,~~~~c_3 \sim 1~,
$$
where for $\langle~ \hat{r}~ \lambda ~\rangle_{ \hat{r}}$ we used
that at the scale $\sim r_0$ the growth of $\lambda$ is correlated
with the fluctuation of $\hat{r}$. Hence we have
 \bel{wiG}
\mathcal{G}(y) ~\simeq~ c_3 ~r_0^2 ~y~~,
 \ee
which agrees with (\ref{delR}).

Note that the form (\ref{borac} - \ref{gapr})~ for
$W[\lambda(y,x)]$ can suggest the analogy with the Luscher's type
oscillation of a string of length $L = 2\pi R$.~ There the quantum
string broadening is $ \sim r_0 \ln L/r_0$. This corresponds to
the \Fd edge width $\sim \ln R(y)$, and this width comes from the
string (\Fd edge) oscillations with the mean wave length $\sim
\sqrt{r_0 L}$ at the zero temperature. At the finite temperature
$t$ there is an additional contribution to the string broadening
$\bar{\lambda} \sim \sqrt{t L/\kappa}$,~ where $\kappa$ is the
string tension, ~and this broadening comes from the large wave
length $\sim L$.  ~In our case (\Fd edge) the effective  $t \sim
r_0^{-1}$ and $\kappa \sim r_0^{-2}$, ~so we come again to the
same result $\bar{\lambda} \sim \sqrt{\bar{R}(y) r_0}$ as above
(\ref{delR} , \ref{wiG}) .

\end{document}